\documentclass{pasa}%

\title[Super-AGB Stars]{Super-AGB Stars and their role as Electron Capture Supernova progenitors} 
\author[Doherty et al.]{Carolyn L. Doherty$^{1,2}$, Pilar Gil-Pons$^{3,4}$ Lionel Siess$^{5}$, \and John C. Lattanzio$^2$ \\
\affil{$^1$Konkoly Observatory, Hungarian Academy of Sciences, 1121 Budapest }%
\affil{$^2$Monash Centre for Astrophysics, School of Physics and Astronomy, Monash University, Australia}%
\affil{$^3$Polytechnical University of Catalonia, Barcelona, Spain}%
\affil{$^4$Institut d'Estudis Espacials de Catalunya, Barcelona, Spain}
\affil{$^5$Institut d'Astronomie et d'Astrophysique, Universit\'e Libre de Bruxelles, ULB, Belgium}}%

\jid{PASA}
\doi{10.1017/pas.\the\year.xxx}
\jyear{\the\year}

\def\msun{M$_\odot$}
\def\lsun{L$_\odot$}
\def\rsun{R$_\odot$}
\def\myr{M$_\odot$\,yr$^{-1}$}
\newcommand{\chem}[1]{$^{#1}$}
\newcommand{\iso}[1]{$^{#1}$}

\usepackage{aas_macros}
\usepackage{hyperref} 
\hypersetup{colorlinks,citecolor=blue,linkcolor=blue,urlcolor=blue}

\usepackage{natbib}
\usepackage{wrapfig}
\usepackage{enumerate}  
\usepackage{mathptmx}

\usepackage[normalem]{ulem}

\begin{document}%
\begin{abstract}

We review the lives, deaths and nucleosynthetic signatures of 
intermediate mass stars in the range $\approx 6.5\,$--$\,12$ \msun, 
which form super-AGB stars near the end of their lives. 
We examine the critical mass boundaries both between different types of massive white dwarfs (CO, CO-Ne, ONe) and between white dwarfs and supernovae and discuss the relative fraction of super-AGB stars that end life as either an  ONe white dwarf or as a neutron star (or an ONeFe white dwarf), after undergoing an electron capture supernova. We also discuss the contribution of the other potential single-star channels to electron-capture supernovae, that of the failed massive stars. We describe the factors that influence these different final fates and mass limits, such as composition, the efficiency of convection, rotation, nuclear reaction rates, mass loss rates, and third dredge-up efficiency. We stress the importance of the binary evolution channels for producing electron-capture supernovae. 
We discuss recent nucleosynthesis calculations and elemental yield results and present a new set of s-process heavy element yield predictions. We assess the contribution from super-AGB star nucleosynthesis in a Galactic perspective, and consider the (super-)AGB scenario in the context of the multiple stellar populations seen in globular clusters. A brief summary
of recent works on dust production is included. Lastly we conclude with a discussion of the observational constraints and potential future advances for study into these stars on the low mass/high mass star boundary .

\end{abstract}
\begin{keywords}
stars: AGB and post-AGB -- stars: evolution -- supernovae: general -- white dwarfs-- nuclear reactions, nucleosynthesis, abundances
\end{keywords}
\maketitle%

\section{Introduction}\label{sec:intro}

Stars in the mass range $\approx 6.5\,$--$\,12$ \msun{} bridge the divide between 
high mass stars and low mass stars, and evolve through a  
super asymptotic giant branch (super-AGB) phase characterised by 
degenerate off-centre carbon ignition prior to the thermally pulsing phase.  
While super-AGB models for the first few thermal pulses have existed for 
quite some time \citep[e.g.][]{pap1,pap2}, it is only relatively recently that there has been a resurgence in their study and that full evolutionary models have been computed for the entire thermally 
pulsing  phase 
\citep[e.g.][]{siess2010,ventura2011a,lau2012,karakas2012,gilpons2013,ventura2013,jones2013,doherty2015}.
Two major reasons why this class of star 
had remained relatively understudied for so 
long are the computational difficulties of following 
degenerate off-centre carbon 
ignition and the very large number of thermal pulses 
expected for super-AGB stars, 
ranging from tens to even thousands. 

One important and highly desirable outcome from stellar calculations for this mass range is a
determination of the final fate of such objects.  The three critical  masses\footnote{These masses are often given different names in the literature; $M_{\rm{up}}$ is also known as $M_{\rm{CO}}$, $M_{\rm{n}}$ is also know as  $M_{\rm{EC}}$ and $M_{\rm{mas}}$ is also known as $M_{\rm{up}}$, $M_{\rm{up^{\prime}}}$, $M_{\rm{up^*}}$, $M_{\rm{min}}$, $M_{\rm{W}}$, $M_{\rm{mass}}$ $M_{\rm{crit}}$ and $M_{\rm{ccsn}}$.} for intermediate mass stars, each of which depends on the stellar composition are: 

\begin{enumerate}
    \item $M_{\rm{up}}$, the minimum mass required to ignite carbon;
    \item $M_{\rm{n}}$, the minimum mass for creation of a neutron star;
    \item $M_{\rm{mas}}$, the minimum mass defining the regime of massive stars, specifically
    those which undergo all stages of nuclear burning and explode as iron core collapse supernovae
    (FeCC-SNe)\footnote{We use SN for ``supernova'' and SNe for the plural ``supernovae''.}. 
\end{enumerate}  
In the standard picture, a star with a mass below $M_{\rm{up}}$ will end its life as a CO 
white dwarf (WD). Stars with masses between $M_{\rm{up}}$ and $M_{\rm{n}}$ leave 
either a CO-Ne or ONe WD remnant, whilst stars with masses between $M_{\rm{n}}$ and
$M_{\rm{mas}}$ undergo an electron-capture supernova (EC-SN), 
ending their lives as neutron stars
\citep[e.g.][]{nomoto1984,ritossa1999,siess2007,poelarends2008,jones2013,doherty2015}. 

A considerable amount of study had been devoted to the explosive deaths of stars in mass range $8\,$--$\,12$ \msun{}, in particular their potential demise as EC-SNe. The earliest works \cite[e.g.][]{miy80,nomoto1984,hil84,nomoto1987} involved the evolution of ``helium balls'' with core masses $\sim 2\,$--$\,2.6$ \msun{} through He and C burning with the
resultant ONe cores then evolved to conditions very close to the expected explosion. Electron capture SNe are caused by the reduction of pressure support due to electron capture reactions on \chem{24}Mg and \chem{20}Ne in stars with H-exhausted core masses $\sim$ 1.375 M$_\odot$ \cite[]{miy80,hil84,nomoto1987}\footnote{This value can vary slightly between calculations, with a slightly lower value of 1.367 \msun found by \cite{takahashi2013}}. The electron captures on these isotopes lead to a reduction 
both in electron fraction ($Y_e$) and the Chandrasekhar mass, 
which triggers contraction \citep{miy80,nomoto1987}. 
Within this collapsing core, the competition between the energy release by O burning and the reduction in electron pressure due to electron capture reactions determines the fate of the ONe core.  
Oxygen is ignited centrally and forms a deflagration that burns the central regions into nuclear statistical equilibrium. The electron capture reactions on this equilibrated material work to further reduce the central density and the subsequent rapid contraction leads to a core collapse. Due to their formation history, EC-SNe are expected to receive a low natal kick \cite[][]{podsi2004,van07,wanajo2011}, have a low explosion energy and small \chem{56}Ni production \citep{kitaura2006,wanajo2017}.

The study of \cite{ritossa1999} was the first to follow the stellar structure of a thermally pulsing super-AGB star, including the stellar envelope, to conditions close to collapse. In recent years a new generation of progenitor models of EC-SNe has been computed which now evolve super-AGB stars to conditions of collapse \citep{takahashi2013} including along the entire thermally pulsing super-AGB phase  \citep{jones2013}.

Determining the mass boundary between stars that do and do not
explode as supernovae 
is a topic of vital importance in astrophysics, for many reasons.
For example, the supernova rate in part determines the number of neutron 
stars and the total energy 
released by supernovae into the environment. 
Based on a standard initial mass function (IMF), there are as many stars born with masses between 5 and 10 \msun{} 
as there are with masses greater than 10 \msun, so how these elusive stars live and die is of interest to many subfields of astrophysics.
This mass boundary is also important for galactic chemical evolution and dust evolution models because stars on either 
side of this divide have significantly different chemical and dust production properties.
Due to the shape of the IMF, super-AGB stars are
both the rarest of the low/intermediate mass stars, and also the most common of the stars 
on the more massive side of the boundary. Hence if they
do indeed produce EC-SNe then they may make a significant 
contribution to the overall SN rate.

Until recently there have been few chemical yields available for these stars.
Chemical evolution 
calculations had to use some strategy to deal with missing yields for
this mass range. The two most common strategies were to either
totally ignore  the yields for this mass range or interpolate in mass
between yields for the low and high mass stars. Either is likely to 
introduce significant errors.

The evolution of massive AGB stars (at the low mass end) and massive 
stars (at the high mass end) is very different and the evolution between 
these is qualitatively different to both, so interpolation is very 
unlikely to be close to the actual yields. Because reliable
yields have been missing, super-AGB stars have long been suspected to 
contribute to solving various astrophysical problems, 
such as the origin of the multiple populations in globular clusters.
We return to this question later.

Super-AGB stars are very difficult to identify observationally, 
with no confirmed detections extant. There is only one 
strong candidate, the very long period (1749 days) and high luminosity ($M_{\rm bol} \simeq -8.0$) star 
MSX SMC 055 \citep{Groenewegen2009}.  Another hindrance to identifying super-AGB stars 
is that their high luminosities and very large, cool, red stellar envelopes 
make them almost indistinguishable from their slightly more massive 
red super-giant counterparts. Indirect evidence for super-AGB stars comes 
from observations of massive O-rich white dwarfs \citep{gan10}, 
and also from neon novae \citep{jos98,wan99,dow13}, with the neon from 
which their name derives assumed to have been dredged-up from the 
interior of ONe WDs, the remains of an earlier super-AGB phase.

In Section 2 we discuss the main evolution characteristics of intermediate mass stars 
including the thermally pulsing super-AGB phase.  In Section 3 we examine the mass limits 
defining the various evolutionary channels, in particular the final fates 
of super-AGB stars, including the importance of the binary star channel 
for formation of EC-SNe.
In Section 4 we describe the nucleosynthesis and stellar yields 
from super-AGB stars. We apply these yields to the globular cluster 
abundance anomaly problems and examine their relative galactic contribution, and briefly touch upon dust production by super-AGB stars. 
Lastly in Section 5 we discuss the observational studies, reiterate the most critical uncertainties 
and discuss future directions in super-AGB star research.
In this review we do not consider super-AGB stars in the early universe. 
For an review on the evolution of primordial and extremely 
metal-poor super-AGB stars we refer to the companion paper
by Gil-Pons (2017) in this edition.

\section{Evolution}\label{sec:evolution}

\subsection{Phases prior to carbon burning}\label{sub:preC}

\begin{figure}
\begin{center}
  \includegraphics[width=8.4cm]{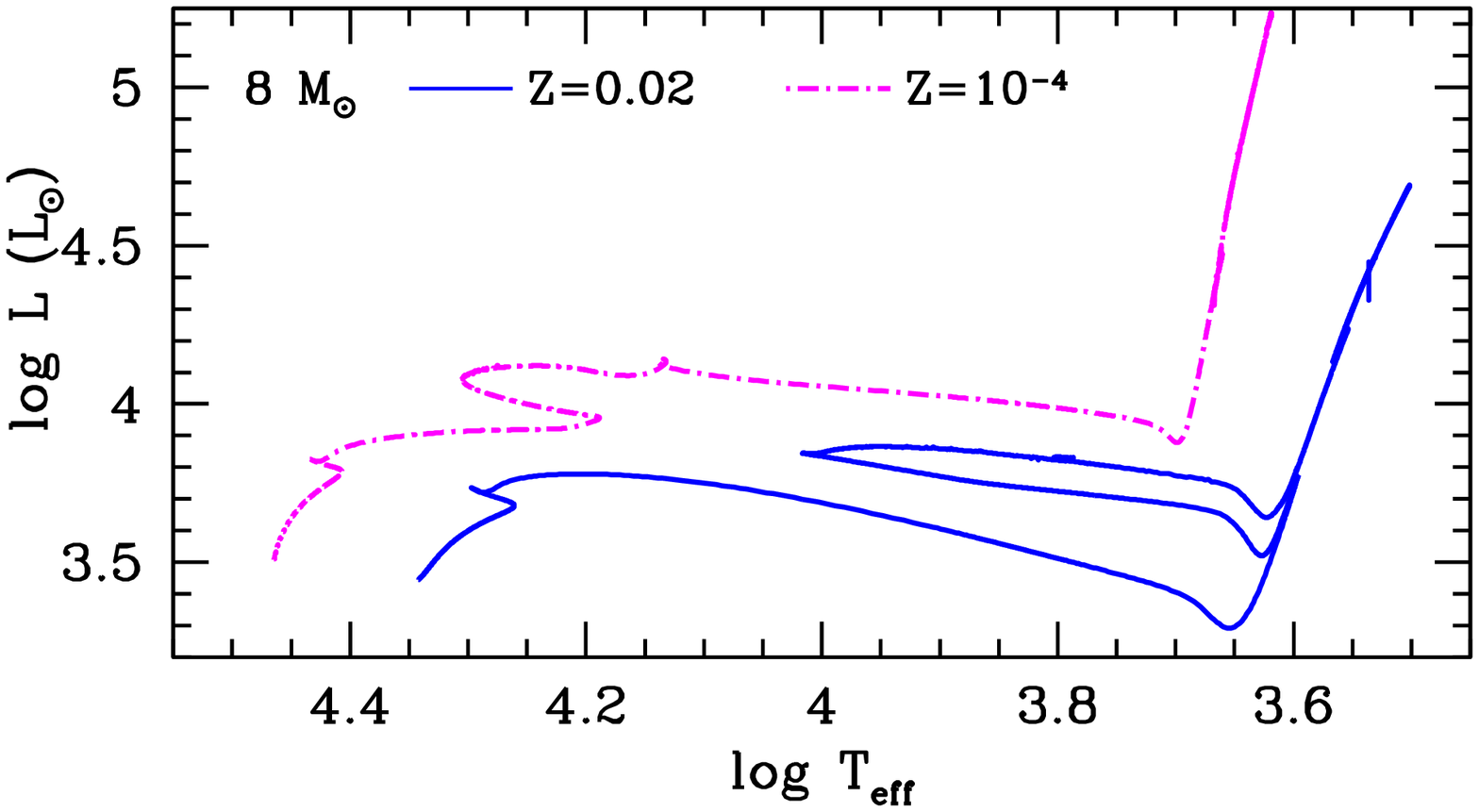}
\includegraphics[width=8.4cm]{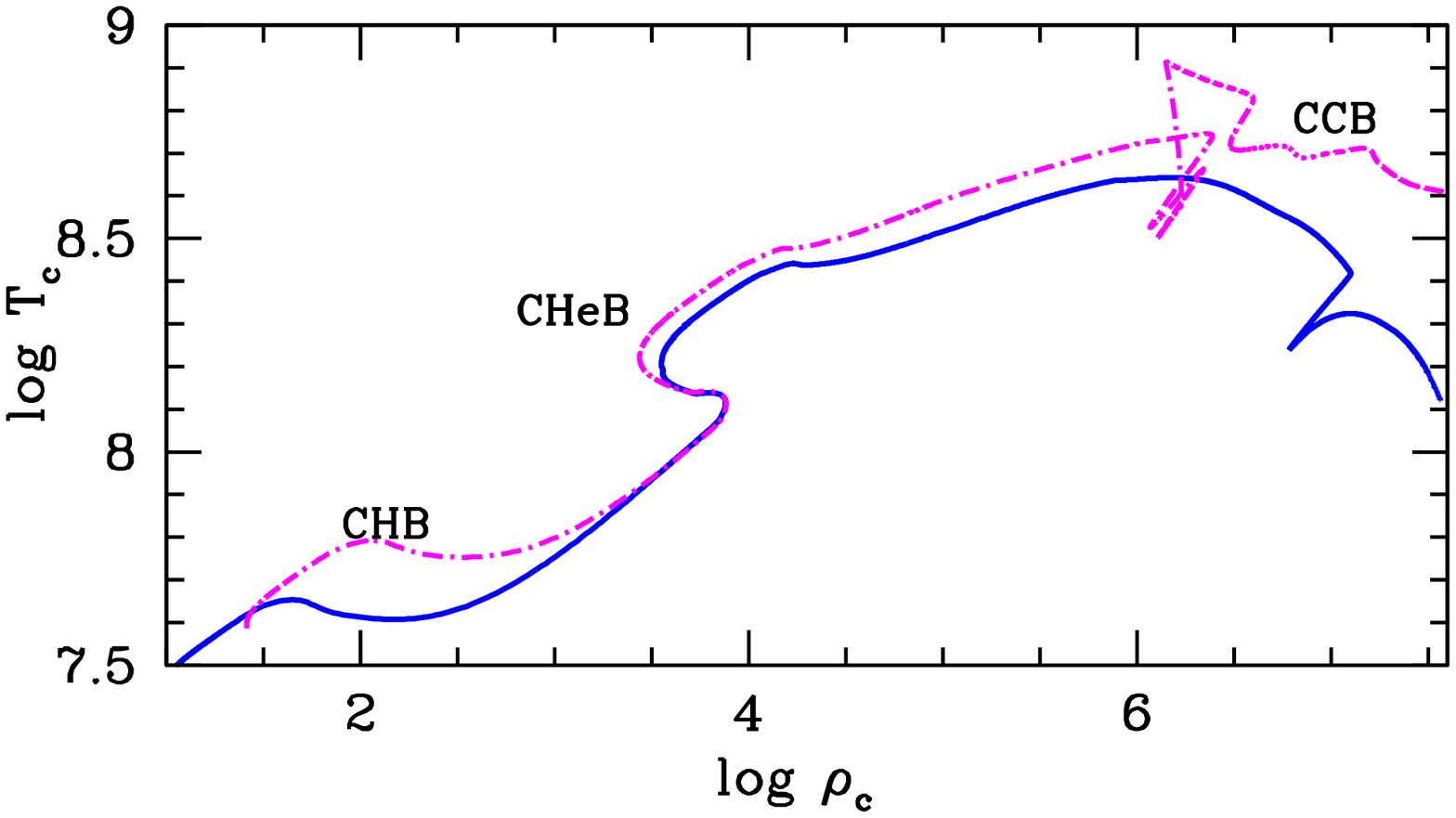}
\caption{Evolution in the Hertzsprung-Russell diagram (top panel) and in the log central density versus log central temperature diagram (bottom panel) of the 8 M$_\odot$ models of super-AGB stars of metallicities Z=0.02 and 10$^{-4}$ from \cite{doherty2015}. CHB, CHeB and CCB refer to central H, He and C burning, respectively}\label{fig-hrd}
\end{center}
\end{figure}

The main nuclear burning stages of intermediate-mass stars are 
well known, with the stars undergoing convective core H-burning (CHB) 
via the CNO cycles followed by convective core He-burning (CHeB).
In Figure~\ref{fig-hrd} we show the Hertzsprung-Russell diagram for 
two 8\msun{}  models of metallicities $Z=0.02$ and 10$^{-4}$ through to the super-AGB phase.
Clearly seen in this figure is the impact of metallicity on the evolution, with the lower metallicity object both more luminous and hotter for the same initial mass.
At decreasing metallicity, stars attain higher central temperatures to counteract fewer CNO seeds (and associated energy generation). These factors result in a larger core mass for the same initial mass. Due to the more rapid ignition of CHeB in the lower metallicity model the (first) 
giant branch is avoided and the star does not undergo a first dredge-up event \citep{gir96}.  
Figure~\ref{fig-hrd} also shows the evolution of central temperature $T_{\rm c}$ versus central density $\rho_{\rm c}$ for the models previously described. We clearly see the occurrence of central H burning 
at higher temperatures 
for the lower metallicity model. Once central H is exhausted, the evolution of the stars in the  $T_{\rm c}$--$\rho_{\rm
c}$ diagram becomes very similar because of the strong dependence of the nuclear burning rates on temperature.  

During CHeB the core is converted to \iso{12}C and \iso{16}O via 
the triple-$\alpha$ and \iso{12}C($\alpha,\gamma$)\iso{16}O reactions with 
the \iso{14}N produced from previous CNO cycling being converted to \iso{22}Ne via 
the reaction chain 
\iso{14}N($\alpha,\gamma$)\iso{18}F($\beta^+\nu$)\iso{18}O($\alpha,\gamma$)\iso{22}Ne. 
The central \iso{12}C content varies strongly with the core mass, 
with more massive cores having less residual carbon due to higher internal 
temperatures \citep{siess2007}. Typically, intermediate mass stars have 
carbon mass fractions of $\sim 0.2\,$--$\,0.5$ at the end of CHeB, with this value highly dependent
on the \iso{12}C($\alpha$,$\gamma$)\iso{16}O reaction rate and treatment of mixing \citep{imbriani2001,str03}.

The duration of the CHB and CHeB phases varies with metallicity and a 
variety of other factors such as treatment of convective boundaries,  semiconvection, rotation\footnote{The main sequence lifetime of intermediate mass stars is well know to increase due to rotational mixing with an approx 20-25$\%$ increase found by \cite{eks12}. However differences in this lifetime between rotating and non-rotating models are lower at only about 5$\%$ in the recent MIST calculations \citep{mist} which use a less efficient rotational mixing.} and nuclear reaction rates,  particularly  the rate of the  
\iso{12}C($\alpha$,$\gamma$)\iso{16}O reaction during the later phases of CHeB. 
Typically the main sequence lifetime of intermediate mass stars in the mass range 
considered here ($\approx 6.5\,$--$\,11$ \msun) is between about $18\,$--$\,60$ Myr with the 
CHeB phase being considerably shorter, of the order $2\,$--$\,5$ Myr.
For the same initial mass, intermediate-mass stars of lower metallicity will have a shorter lifetime albeit only slightly, with the 8\msun{} models presented here showing a difference in main sequence lifetime of only $\sim$ 10$\%$.

\subsection{Carbon burning phase}\label{sub:cburn}

Very detailed descriptions of the carbon burning phase within super-AGB stars can be found in 
works such as \cite{pap1}, \cite{siess2006} and \cite{farmer2015} whilst here we provide a brief overview.

Once CHeB has ceased, the resulting CO core 
begins to contract\footnote{The contraction time from the cessation of CHeB to C ignition is a function of core mass with larger core masses evolving more rapidly, with typical values of $1.5$--$3 \times$ 10$^5$ yr \citep{doherty2010}.} causing an increase in the central density 
which leads to neutrino energy losses becoming important for the innermost regions of the star.
This results in cooling and the formation of a temperature inversion. When the 
peak temperature reaches approximately 640MK, and the density at that point
is about 1.6 $\times$ 10$^6$ g cm$^{-3}$, carbon is ignited. 
This ignition takes place off-centre and under conditions of partial 
degeneracy ($\eta$ $\sim$2-3). The peak carbon burning luminosity 
during this initial violent carbon burning flash can be up to about 10$^9$~L$_\odot$ and the
large energy release drives the formation of a convective zone. 
After a short period this first carbon flash is quenched and contraction of 
the core resumes, to be followed by another carbon flash. In this second flash 
the degeneracy is lower ($\eta$ $\sim$1) and the convective region that forms 
(classified as a ``flame''') subsequently burns inwards until it reaches the centre. 
Carbon burning however is not complete and continues radiatively outwards, 
generating secondary convective flashes when regions of high carbon content 
are encountered. The specific  number of 
flashes depends on the degeneracy of the core,
and thus on the star's initial mass. 
In general lower mass objects tend to experience a higher number of secondary flashes and with higher intensity. 

The effect of the carbon burning flashes and flame on the central region can be seen in the bottom panel of
Figure~\ref{fig-hrd}. For the Z=$10^{-4}$ model the initial flash causes a sharp drop in both temperature   
and density (down to log~T$\sim 8.5$ 
and log~$\rho \sim 6$). The point at which the flame reaches the centre 
is quite evident, being characterised by the large and steep rise in the central 
temperature at about constant density (log~$\rho \sim 6$) \footnote{For the Z=0.02 model there is 
only one flash and then the core cools without any further C burning. 
This star forms a hybrid CO-Ne white dwarf (Section.~\ref{sub:cone}).}.
The strength of the initial carbon flash is larger in the more degenerate 
(i.e. less massive) models, with the carbon flash luminosities ranging 
from $\sim$ 10$^6$ to 10$^9$ L$_\odot$. The more massive models also ignite carbon closer to the centre, 
in conditions of milder degeneracy than the lower mass cases. 

\begin{figure}
\begin{center}
\includegraphics[width=8.4cm,angle=0]{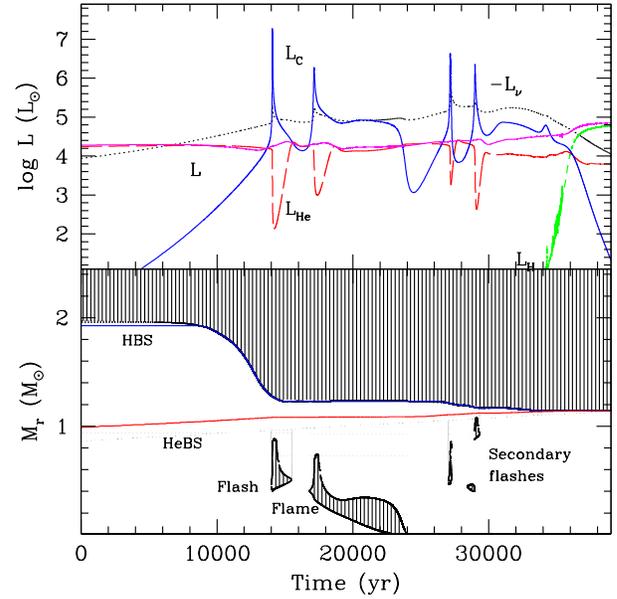}
\caption{Kippenhahn and luminosity diagram during the carbon 
burning phase for an $8.5$\,\msun{} model 
with $Z=0.02$ from \cite{doherty2015}. Time has been 
set to zero when $L_{\rm C}$ first exceeds  $1$\,\lsun. 
In the upper panel we show
different luminosity sources:  H in green, He in dashed red,
C in blue, surface in magenta, and the negative of the neutrino luminosity is in black. In the 
lower panel the mass coordinate of the HBS is shown in blue, the HeBS in red, and the hatched regions represent convection.}
\label{fig:cburning}\end{center}
\end{figure}

Figure~\ref{fig:cburning} is a Kippenhahn diagram of an 8.5 \msun{} Z=0.02 model and illustrates the typical multi-step burning process, 
consisting of an off-centre flash, a flame that 
propagates towards the centre, then subsequent secondary carbon flashes in the 
outer parts of the core.  In the top panel the evolution of the 
H, He, C, neutrino and total luminosities is shown. Clearly seen are the 
carbon burning flashes/flame with peaks in $L_{\rm C}$. The total luminosity of 
the star is almost constant through the carbon burning phase with its 
behaviour decoupled from the central burning regions. During the steady 
carbon burning flame phase, all of the energy released by carbon burning
is carried away by neutrinos,  in what is called the ``balanced power 
condition'' \citep{timmes1994}, as seen in Figure~\ref{fig:cburning}. 
The carbon burning flame speed is quite slow $\sim$ 10$^{-2} \,$--$\, 10^{-3}$ cm s$^{-1}$ \citep{timmes1994,pap2,siess2006}.
The duration of the carbon burning phase decreases with 
increasing core mass (i.e. initial mass) and ranges between 
about 10,000 and 40,000 years.

The minimum core mass for C-burning occurs for stars which have CO core masses $\gtrsim$ 1.05 $M_{\odot}$ at the start of carbon burning.  
  
The main nuclear reactions during the carbon burning phase are 
\iso{12}C(\iso{12}C,p)\iso{23}Na and \iso{12}C(\iso{12}C,$\alpha$)\iso{20}Ne, followed by 
\iso{23}Na(p,$\alpha$)\iso{20}Ne and \iso{16}O($\alpha$,$\gamma$)\iso{20}Ne.
 The carbon burning rates are quite uncertain and it has been suggested 
 that there may be unmeasured resonances \citep{spillane2007} or hindrances \citep{jiang2007} which may alter the rates by more than a factor of 1000 compared to the standard rates from \cite{cf88}. Due to their importance for a variety of stellar environments, in particular in Type 1a SN studies, these reaction rates are currently under much investigation \cite[e.g.][]{buc15}.  
 \cite{chen2014} examined the impact of variations to the \iso{12}C+\iso{12}C rates on carbon burning within super-AGB stars and found that if the rates were multiplied by factors of 1000 and 0.01 the minimum CO core mass for carbon ignition became 0.93 \msun{} and 1.10\msun{} respectively. 

After the completion of CCB the core has been converted to 
mostly \iso{16}O ($50\,$--$\,70$\%), \iso{20}Ne ($15\,$--$\,35$\%) and trace 
amounts of \iso{23}Na, \iso{24,25,26}Mg, \iso{21,22}Ne and \iso{27}Al \citep{siess2007}.
The third most common element in the cores varies between calculations and is either Mg \citep{nomoto1984,miy80,takahashi2013} or Na \citep{pap1,siess2006} which results in either ONeMg or ONeNa cores\footnote{The amount of Mg or Na may have important implications in the subsequent evolution if the stellar core grows to conditions for an EC-SN e.g \cite{gut05}.}.
There is also a small abundance of \iso{12}C remaining throughout the ONe core of about $0.2\,$--$\,2$\%, with this residual carbon abundance being lower in the more massive models.

In the traditional picture, stars with ONe core masses exceeding 1.37 \msun{}\footnote{This value for neon ignition for a pure Ne core was found to be slightly lower at 1.35 \msun by \cite{schwab2016} their Figure C2.} at the end of C-burning will ignite neon and undergo all stages of further burning \citep{nomoto1984}. Therefore ONe cores are expected to be produced with masses $\sim 1.05\,$--$\,1.37 M_{\odot}$.

\subsubsection{Incomplete C-burning - Hybrid CO-Ne cores}\label{sub:cone}

In stars with initial mass slightly above $M_\mathrm{up}$ carbon ignites in the 
very outer regions of the core. In some cases, after the primary flash has occurred, 
no further carbon burning takes place. These aborted carbon ignition models 
have an interior comprised of a very large central CO region surrounded by 
a thin ONe layer and a further outer CO region \cite[e.g][]{doherty2010,ventura2011a}. 
The initial mass range for creation of this class of hybrid CO-Ne cores is 
very narrow, being at most about 0.1\msun. CO-Ne cores can also be formed 
when the carbon burning flame stalls on its journey towards the centre. 

Due to off-centre (convective) carbon burning in super-AGB stars a molecular weight inversion is created which can drive thermohaline mixing \citep{siess2009}.
The resultant mixing transports carbon from the inner region toward the burning flame, replacing it with the heavier products of carbon burning. Thermohaline mixing below the C-convective shell is thus able to decrease significantly the C content in the zones ahead of the C-burning flame, and deprived of fuel, this causes the extinction of the flame before it reaches the centre.
The result is a hybrid degenerate core, comprising an inner zone of unburnt (but depleted) CO  and an outer ONe zone.
In such a case, in contrast to  evolution ignoring thermohaline mixing, the ONe core 
is left with a larger amount of unburned \iso{12}C, between $2\,$--$\,5$\%, in the centre.
Later using downward revised values of the thermohaline mixing coefficient 
based on multidimensional hydrodynamic simulations, \cite{denissenkov2013} discounted 
the ability of thermohaline mixing to stall the carbon burning flame, 
suggesting that the mechanism was too inefficient. 
They did however suggest another mechanism that may be active during the carbon burning phase within 
super-AGB stars and which could work to halt flame propagation. It is well known that convective 
flows may lead to mixing beyond the strict Schwarzschild boundary, often known as 
convective boundary mixing (CBM). This was applied by \cite{denissenkov2013} to the base of the C-burning convective shell of super-AGB stars. These authors used their results from hydrodynamical 
simulations to simulate the propagation of the C-burning flame and 
determined that it was actually deprived of fuel, and that the C-burning 
process was halted and hybrid CO-Ne cores were formed.
For determining the composition of stellar material, the main difference between thermohaline 
mixing and CBM concerns the extent of the induced mixing. Thermohaline mixing connects the entire interior to the
region with the higher molecular weight that is being produced by the burning flame.
With overshoot the mixing only occurs directly below the convection over a region
whose width is determined by some model or algorithm. 
In contrast to thermohaline mixing, 
this leaves a pristine CO core interior to the maximum extent of the overshoot.  Thus CBM also finds 
hybrid CO-Ne cores, but these hybrids can be produced with a range of 
configurations and quite widely varying widths of the ONe shell. This is unlike 
the structures produced from the lower mass super-AGB stars which 
have thin ONe shells in the far outer core.

\cite{chen2014} investigated the effects of CBM and the quite 
uncertain \iso{12}C reaction rates on hybrid core creation. 
They found that varying the efficiency of CBM in addition to the \iso{12}C reaction rates resulted in
the formation of  hybrid CO-Ne cores over a wide range of core masses from $\approx 0.93\,$--$\,1.30$ \msun. This corresponds to an initial mass range for hybrid CO-Ne core creation (defined as 
$\Delta M_{\rm{CO-Ne}}$) of up to 1 \msun{} which would make the CO-Ne cores very common. This very large core mass of 1.30 \msun~for hybrid CO-Ne WDs could have important implications for the rate of Type 1a SNe \cite[e.g.][]{meng2014,wang2014,liu15}.

\cite{farmer2015} studied carbon ignition within intermediate/massive stars with an extensive grid of models looking at the effects of rotation, convective overshooting\footnote{This followed the work of \cite{her97} by using a diffusion co-efficient $D_{OV}$ beyond the formal convective border where $D_{OV} = D_{0} exp \left(\frac{-2z}{f_{\rm{over}} H_{p}}\right)$ where D$_{0}$ is the diffusion coefficient near the convective boundary, $z$ is the radial distance from the edge of the convective zone, and $H_{p}$ is the pressure scale height at the convective edge. They examined $f_{\rm{over}}$ in the range $0\,$--$\,0.02$.}, thermohaline mixing and combinations of these processes.  In their study they found that a substantial number of stars which ignited carbon off-centre went on to form CO-Ne cores. In particular, in agreement with \cite{denissenkov2013} and \cite{chen2014}, models with efficient overshooting at the base of the convective carbon burning region led to a very wide initial mass range for hybrid CO-Ne WDs.

Recently, work by \cite{lec16} using 3D hydrodynamic simulations has 
suggested that convective mixing cannot stall the carbon burning flame due 
to the large buoyancy barrier that needs to be crossed to reach the radiative burning front 
and hence formation of CO-Ne WDs would not be typical.  Irrespective of whether CO-Ne cores could actually form, \cite{brooks2016} showed that a structure composed of a higher density ONe mantle above a CO core would be unstable to rapid mixing shortly 
after the onset of the WD cooling sequence. Thus the actual occurrence of hybrid cores and their possibility to remain unmixed throughout the latest stages of stellar lives is still a matter of debate.
 
One of the main interests in hybrid cores is related to the potential eventual fates of
SNIa. The amount of available C would probably be high enough so that, 
if the degenerate core were able to increase in mass up to $M_{\rm Ch}$, 
a thermonuclear (single) SN explosion would result \citep{poelarends2008}. 
Alternatively, if the super-AGB star were the primary component of a close 
binary system with specific
initial orbital parameters, SNIa explosions might occur. 

This possibility was explored by \cite{bravo2016}, who computed the 
hydrodynamical explosion of white dwarfs hosting hybrid cores, under 
different conditions (size of the hybrid cores, and ignition by deflagration or detonation). 
These authors showed that SNIa harboring hybrid cores would be characterised by 
lower kinetic energies and lower amounts of ejected \chem{56}Ni than their 
pure CO WD counterparts. Explosions of these hybrid cores may 
be the theoretical counterparts of the sub-luminous 
class of SN2002 cx-like SN or SNIax. \cite{denissenkov2015} also pointed out 
the fact that hybrid CO-Ne cores might be one possible reason 
for the inhomogeneity of observed SNIa. More recent multidimensional 
hydrodynamical simulations by \cite{willcox2016} reproduce the same trend, 
that is, their SNIa models hosting hybrid degenerate cores also produce 
less \chem{56}Ni and release less kinetic energy.

\subsubsection{Incomplete Ne-burning - Failed massive stars}\label{fms}

The ability of the Ne burning flame to propagate to the centre is of crucial importance in determining if the star ends it life as an EC-SN or an FeCC-SN.
As mentioned in Section.~\ref{sub:cburn} if the ONe core mass 
exceeds 1.37 \msun{} it is assumed that Ne shall ignite and 
the star will follow the massive star channel. 
However there are slight complications to this standard picture.
The behavior of Ne burning is very similar to that seen during the 
earlier phase of C burning.
Efficient neutrino cooling causes the temperature maximum to move 
away from the centre, resulting in Ne ignition occurring further 
off-centre for lower masses.
Akin to the aborted carbon ignition models described in Section~\ref{sub:cone}, if neon is ignited at the very outer edge of the core there will be a brief neon flash but no subsequent burning nor flame propagation \citep{timmes1994,ritossa1999,eldridge2004}.  It is expected that these stars with core masses so close to the Chandrasekhar mass will end life as EC-SNe after a very brief thermally pulsing phase. \cite{doherty2015} proposed a new nomenclature for models which undergo only very slight off-centre neon burning and then later reach the TP-(S)AGB phase, calling them ``hyper-AGB'' stars. 

In addition to the super-AGB evolution towards an EC-SN a second possible single star EC-SN channel exists, that of ``failed massive stars'' (FMS) \citep{jones2013,jones2014}. 
A FMS is formed in stars with ONe core masses slightly above the value for Ne ignition. If Ne is ignited far enough off-centre and convective boundary mixing is employed at the base of the Ne burning shell, then instead of a flame progressing smoothly
towards the centre, the Ne burning can be stalled and undergo multiple flashes.  After each flash there is a period of contraction which, given enough time, can ultimately result in the core reaching sufficient densities for the URCA processes to be activated  and the star to subsequently reach conditions for an EC-SN prior to the Ne flame being able to reach the centre.
However, if no CBM is employed and the strict Schwarzschild boundary is used at the base of the Ne convective region, as seen in \cite{jones2014}, then the class of FMS ceases to exist and the Ne flame is free to propagate inward towards the centre with the star most likely becoming an FeCC-SN.
The exact contribution from the FMS to the EC-SN channel is highly uncertain but if this class of star only occur for models in which the H-exhausted core has been reduced to precisely  the Chandrasekhar mass, (refer to next section.~\ref{sub:SDU}) then we expect a narrow channel.

\subsection{Reduction of H-exhausted core mass}\label{sub:SDU}

For intermediate mass stars, the H-exhausted core masses after CHe burning are 
in the range $\approx 1.6\,$--$\,2.6$ \msun.
Hence after CHeB all future super-AGB stars will eventually grow degenerate core masses far exceeding 
the Chandrasekhar mass ($M_{\rm{Ch}}$) and therefore if no process 
takes place to reduce the core mass, these stars will undergo all 
stages of core burning, just as do massive stars. 

\begin{figure}[t]
\includegraphics[width=8.5cm,angle=0]{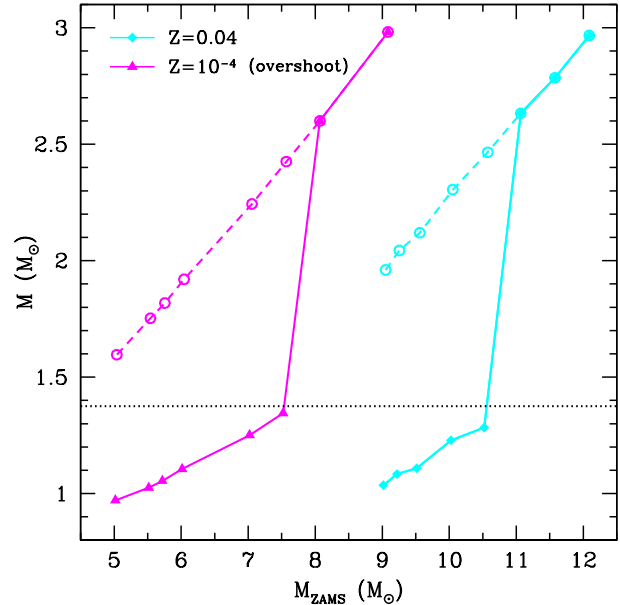}
\caption{Mass of the H-exhausted core before (open circles connected by a 
dashed line) and after (triangles/diamonds connected by a solid line) the operation 
of the SDU for two metallicities. The left/magenta and right/cyan 
lines correspond to models with a metallicity $Z=10^{-4}$ (with core 
overshooting) and $Z=0.04$ (without core overshooting), respectively. The dotted horizontal line represents the Chandrasekhar mass. Models are from \cite{siess2007} with overshoot as described in \cite{her97} with a value $f_{\rm{over}}=0.016$ (see footnote~9).}
\label{fig:SDUp}
\end{figure}

Prior to the thermally pulsing phase two processes can reduce 
this H-exhausted core mass, these being second dredge-up (hereafter SDU) and dredge-out.
Figure~\ref{fig:SDUp} shows the H-exhausted core mass both before and after SDU. Clearly seen in the sharp divide between stars that undergo SDU and those that do not. This figure also highlights that this same behaviour occurs over a large spread in metallicity, and both with and without convective overshooting. 

\subsubsection{Second Dredge-up}

Due to the gravitational contraction of the core after CHeB, the envelope expands and cools, with 
convection penetrating inwards into the H exhausted core \citep{bec79}. 
The SDU event occurs at different stages of the C-burning phase 
for stars of different initial masses. For low-mass super-AGB stars it 
takes place prior the first C-flash. Stars of higher initial mass evolve faster 
and thus ignite C earlier. 
Normally SDU only brings to the surface material that has undergone H burning.
However, the more massive stars can experience what is called a ``corrosive''' SDU 
episode \citep{gilpons2013,doherty2014b}: in these cases the 
bottom of the convective envelope is able not only to reach below the erstwhile
HBS, but also to reach deeper, to where reside the
products of the HeBS. 
Corrosive SDU enriches the surface with substantial amounts of 
primarily \chem{12}C \cite[and \chem{18}O, e.g.][]{bec79,her04a}, 
while in the more massive models corrosive SDU also enriches the surface with
substantial amounts of \iso{16}O. The masses of the 
stellar cores for which corrosive SDU occurs vary between studies, with 
about $1.15 \,$--$\,1.28$ \msun{} in \cite{doherty2014b}, down to 1.03 \msun{} in \cite{her04a}. These values are generally lower for lower metallicity models due to their broader residual He shells. 
The degree of metal enrichment from SDU in the envelope of low metallicity stars is critical for their future evolution. 

\subsubsection{The dredge-out episode}\label{sub:do}

\cite{ritossa1999} first named, described and provided an extensive analysis of the phenomenon known as dredge-out. It was later reported by \cite{siess2007}, \cite{poelarends2008}, \cite{takahashi2013}, \cite{gilpons2013}, \cite{doherty2015} and \cite{jones2016a} using different evolutionary codes and different input physics (in particular the treatment of mixing and treatment of convective borders). This phenomena occurs for massive super-AGB stars regardless of their metallicity \cite[e.g.][]{gilpons2013} and occurs for stars in the upper $\approx$ 0.3 M$_{\odot}$ range of super-AGB stars. 

\begin{figure}
\begin{center}
\resizebox{\hsize}{!}{\includegraphics{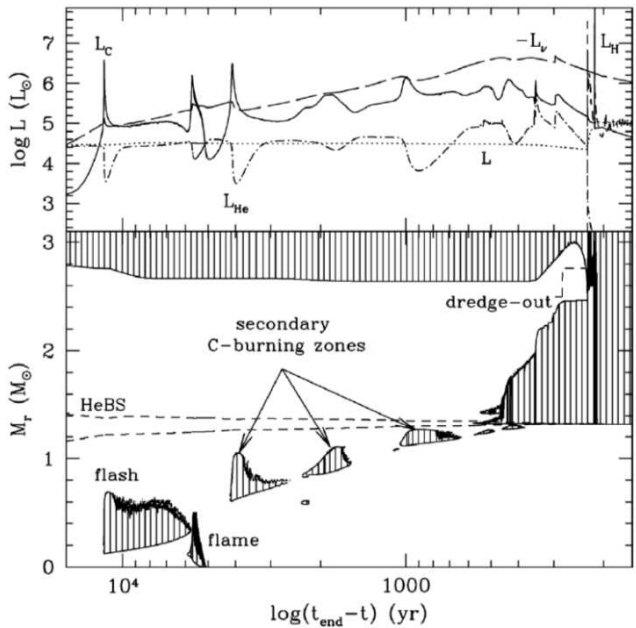}}
\caption{
Kippenhahn and luminosity diagram during the carbon burning phase and dredge-out episode for an $9.5$\,\msun{} $Z=0.001$ model from \cite{siess2007}. Time is counted backwards from the last computed model.}
\label{dout}
\end{center}
\end{figure}

Figure~\ref{dout} shows the evolution during C-burning and dredge-out phase for a 9.5 M$_{\odot}$ Z=0.001 star from \cite{siess2007}.
Nearing the end of the carbon burning phase a convective He shell develops near the upper boundary of the partially degenerate core. This shell is initially separated from the base of the convective envelope by a relatively extended radiative region (about 1\msun) and a thin semiconvective region near the He-H interface. As described in \cite{ritossa1999} the He convective shell is initially sustained mainly by C-burning, and gravothermal energy, but He-burning powers its final approach towards the base of the convective envelope. Eventually these convective regions meet and protons are ingested into very high temperature ($\gtrsim 10^8$ K) He- and C-rich regions. These ingested protons rapidly undergo the \chem{12}C(p,$\gamma$)\chem{13}N reaction leading to a H-flash with peak luminosities of $L_{\rm H} \sim$ 10$^9$ L$_{\odot}$. The associated total energy release from the H-flash is vast, and generated in a very small region.
 According to estimates by \cite{jones2016a} the energy released by this process 
represents about 11$\%$ of the star's internal energy and about 8$\%$ of its binding energy. 
It is likely that this has hydrodynamical consequences and the assumption of hydrostatic equilibrium should be doubted. At the very least, it is likely that time-dependent convection is required \citep{herwig2011}.
 \cite{jones2016a} suggest that this dredge-out may provoke a global oscillation of shell-H ingestion (GOSH) event \citep{herwig2014} which could potentially drive more powerful and non-radial hydrodynamic events leading to mass ejection.

With an abundant supply of \chem{13}C now in a high temperature, helium-rich region, the \chem{13}C($\alpha$,n)\chem{16}O reaction is expected to take place at a rapid pace and produce a substantial number of free neutrons \citep{doherty2015,jones2016a}. 
A dredge-out event is expected to produce neutron densities of the order N$_n$ $\approx 10^{15}$\;cm$^{-3}$ corresponding to the intermediate n-capture regime (known
as the ``i-process'', see \citealp{cowan1977}). This process in super-AGB stars was suggested by \cite{jones2016a} to be responsible for the occurrence of some carbon-enhanced metal poor stars enriched in s- and r-process elements (the CEMP s/r stars, see \citealp{beers2005}). But based on an IMF argument, the relatively few super-AGB stars seem unlikely to be a major source of pollution of the CEMP s/r stars \citep{abate2016}.

Besides the possibility of ejection of heavier-than-iron elements, the dredge-out process also alters surface abundances of light elements, in particular He and the He-burning product \iso{12}C \citep{ritossa1999}. This results in the most massive super-AGB stars becoming carbon stars. 

In models of slightly lower mass than those that undergo dredge-out, near the end of carbon burning and prior to the SDU there is also the formation of a convective He region. However this convective zone decays before the convective envelope penetrates inwards and therefore does not merge with the proton-rich region.  This material will be highly enriched in \iso{12}C \cite[e.g.][]{herwig2012}.

\subsection{TP-super-AGB phase}\label{tpasgb}

\begin{figure*}
\begin{center}
\includegraphics[width=17cm,angle=0]{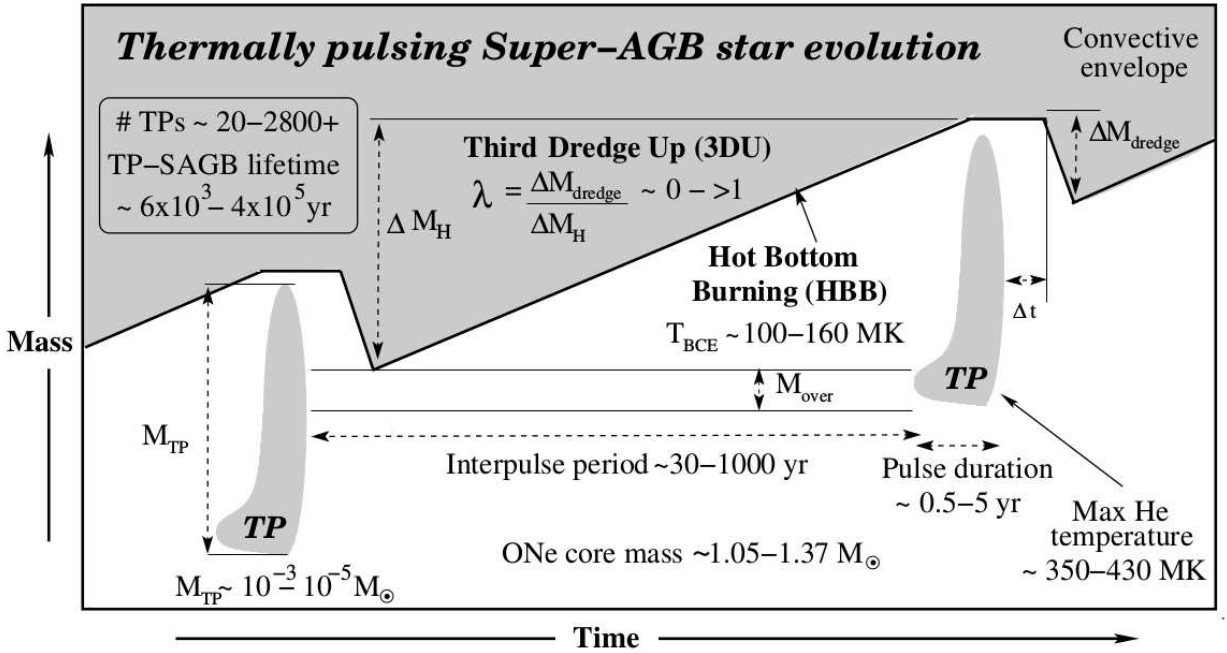}
\caption{Schematic Kippenhahn diagram of two consecutive thermal pulses showing typical values for super-AGB stars. The upper light grey shaded region represents the convective envelope and the two thin shaded regions represent the convective shells associated with two consecutive flashes.}
\label{fig-schematic}
\end{center}
\end{figure*}

After the cessation of core carbon burning a super-AGB star consists of a massive ONe core surrounded by a CO shell, 
a H-burning shell and a very extended H-rich envelope. Quiescent H-burning is eventually interrupted by unstable He-burning, a thermal pulse ensues and the thermally-pulsing super-AGB (TP-SAGB)
phase begins. Early He-flashes tend to be relatively mild, but their peak luminosities grow as the 
evolution progresses. 
When the TP-SAGB is established, H-burning and He-burning in shells alternate as nuclear energy suppliers. 
Figure~\ref{fig-schematic} gives a schematic overview of the typical values from the literature associated with the thermally pulsing phase of super-AGB stars. 

Whilst qualitatively similar to their lower mass counterparts\footnote{For a recent review of AGB stars refer to \cite{kar14}.}, super-AGB stars present some important differences. 
The most obvious is that the stellar cores and envelopes are more massive, between $\approx 1.05\,$--$\,1.37$ \msun, and $\sim 5\,$--$\,10$ \msun{} respectively. 
Due to their larger, hotter and more compact cores the recurrence time between thermal pulses (the interpulse period) is much shorter (10s-1000s yr) in super-AGB stars and due to this they can undergo from between tens to multiple thousands of thermal pulses, typically 
with more pulses at lower metallicity. The thermal pulse duration is also greatly reduced in comparison to lower mass AGB stars, with pulses lasting only about $0.5\,$--$\,5$ yr. The intershell
convective regions are also thinner with the mass region of only 10$^{-3}$ to 10$^{-5}$ \msun.

The maximum temperature within the HeB convective zone steadily increases throughout the evolution along the TP-SAGB and also increases with increasing initial mass, with the most massive super-AGB star models achieving temperatures in the range $350\,$--$\,430\,$MK. This high temperature has important implications for the activation of the \iso{22}Ne neutron source and heavy element production (see Section.~\ref{subsec:heavy}).
 The strength of the thermal pulses, as measured by the HeB 
 luminosity $L_{\rm He}$,  decreases for super-AGB stars with increasing (initial) core mass. This is due to the reduced temperature sensitivity of the triple $\alpha$ reaction, the higher radiation pressure and the lower degree of degeneracy \cite[e.g.][]{sac77,sug78,siess2006}. This $L_{\rm He}$ value varies widely between computations from different research groups and typically those with less violent thermal pulses have lower dredge-up efficiency. 
The overlap factor $r$ is defined by $r  = M_{\rm{over}}/M_{\rm{TP}}$ where $M_{\rm{over}}$ is the mass 
contained in the previous intershell convective zone that is engulfed in the next pulse {(see Figure~\ref{fig-schematic})}
and $M_{\rm{TP}}$ is the mass of the intershell convective zone at the current thermal pulse.  This parameter is important in particular in relation to heavy element production because it 
determines the amount of material that 
experiences multiple neutron exposures in subsequent thermal pulses. 

Due to the activation of nuclear burning at the base of their convective envelopes super-AGB stars are more luminous than the classical AGB limit \citep{pac70}. The most metal poor models can reach in excess of $10^{5} L_\odot$ ($M_\mathrm{bol} \sim -8.2$), which places them at comparable luminosity to the more massive red super-giants.

The third dredge-up (hereafter TDU) is the process where after a thermal pulse the convective envelope penetrates through the (extinct) hydrogen shell and into the intershell region, mixing up products of H and partial He burning. Apart from the enrichment of the envelope composition another important consequence of the TDU is the reduction in the mass of the H exhausted core. 
The efficiency of TDU is commonly measured using the $\lambda$ parameter, defined as $\Delta M_{\rm{dredge}}/\Delta M_{\rm{H}}$ where $\Delta M_{\rm{H}}$ is the increase in the core mass during the previous interpulse phase and $\Delta M_{\rm{dredge}}$ is the depth of the dredge-up (see Figure~\ref{fig-schematic}).  By this definition a $\lambda$ value of one represents the case where the entire region processed by the H shell during the previous interpulse phase is mixed to the surface during the subsequent TDU episode and there is no overall core growth. 
We note that there is no physical reason why $\lambda$ cannot exceed unity. 

In super-AGB star modelling the amount (and even occurrence) of TDU is hotly debated, with computations finding no TDU \cite[e.g.][]{siess2010,ventura2013}, low efficiency TDU with $\lambda\sim 0.07-0.3$ \citep{pap2}, moderate efficiency
TDU with $\lambda\sim 0.4-0.8$ \citep{doherty2015} and high efficiency TDU with $\lambda >$ 1 \citep{herwig2012,jones2016a}.
Although quantitative differences exist between evolutionary calculations from different research groups, the general trend is for decreasing TDU efficiency (or cessation of TDU entirely)  as one transitions from intermediate/massive\footnote{We define massive AGB stars as those with initial masses  $\gtrsim$ 5\,M$_\odot$ but not massive enough to ignite carbon.} AGB stars to super-AGB stars (and larger core masses).

\cite{jones2016a} examined the variations to the efficiency of TDU caused by modifying the amount of convective boundary mixing at both the base of the intershell convective zone and the convective envelope. With significant convective boundary mixing included  they uncovered a new potential convective-reactive site, where the TDU begins whilst the convective thermal pulse was still activated (i.e. $\Delta t < 0$ in Figure~\ref{fig-schematic}), leading to an ingestion of protons within the convective 
thermal pulse. We return to the potential nucleosynthesis implications of these unusual thermal pulses in section~\ref{subsec:heavy} 

Super-AGB stars are very luminous, cool objects (2500 K $<$ T$_{\rm{eff}}$ $<$ 4000 K) with large distended envelopes of $R\gtrsim$ 1000 \rsun. During their lives they can lose a substantial amount of material, up to about 90$\%$ of their initial mass, through stellar winds. During the majority of the TP-SAGB phase the mass loss is in the superwind phase, with mass loss rates up to $\sim$ 10$^{-4}$ \msun yr$^{-1}$, 
and expansion velocities up to about 25 km s$^{-1}$. 
When commonly used AGB star mass loss rates such as those by \cite{vassiliadis1993} or \cite{bloecker1995} are applied to super-AGB stars the average mass loss rate during the TP-(S)AGB phase ranges from $0.1\,$--$\,3\times10^{-4}$ M$_\odot$ yr$^{-1}$. 
Rapid mass loss from super-AGB stars means that
their thermally pulsing lifetimes are quite short, of the order of $10^{4}$ to $10^{5}$ years.

Super-AGB star model computations cease due to convergence problems prior to the removal of the entire envelope. This can occur when the amount of remaining envelope is still quite large, up to $\sim$ 3\,M$_\odot$. The loss of convergence generally occurs just after a thermal pulse when the 
radiation pressure is very high and the contribution of the gas pressure to total pressure tends to zero in shells near the base of the convective envelope. This results in super-Eddington luminosities \citep{wood1986,wagenhuber1994}. This instability has been attributed to the presence of an opacity peak due to iron in these layers of the star \cite[][]{sweigart1999,lau2012}.
This will likely lead to the inflation of the envelope and 
either its entire ejection or a period of enhanced mass loss. Envelope inflation
due to the Fe opacity also occurs in massive stars 
\cite[e.g.][]{petrovic06,grafener11}. We expect that
multi-dimensional hydrodynamics will be required to understand the occurrence and outcome of such events \citep{jiang2015}.

After leaving the thermally pulsing phase, super-AGB stars are expected to go through a short lived planetary nebula (PN) phase before reaching the white dwarf cooling track.


\section{The Mass Range of super-AGB Stars}\label{sec:masses}

The precise lower and upper initial mass limits for stars that will enter the super-AGB phase 
depend on the input physics and on numerical aspects of the calculations. As mentioned in the introduction there are three important mass limits in the intermediate mass regime: $M_{\rm{up}}$, $M_{\rm{n}}$ and $M_{\rm{mas}}$. The difference between the $M_{\rm{up}}$ and $M_{\rm{mas}}$ values sets the maximum (initial) mass range for super-AGB stars. In the following subsection we will examine how these boundary values change with differing compositions and mixing approaches and also discuss complications to this standard picture. 

\subsection{The Critical Masses $M_{\rm{up}}$  and $M_{\rm{mass  }}$}\label{critmass}

Figure~\ref{fig-mupmmass} is a compilation of M$_{\rm up}$ (bottom panel) and M$_{\rm mas}$ (top panel) values from the literature and illustrates both the large spread in results between different research groups and also the behavior of these quantities with initial metallicity.

The mass boundaries $M_{\rm{up}}$ and $M_{\rm{mas}}$ are highly dependent on the maximum convective core mass obtained during CHB and CHeB. As seen in section~\ref{sub:cburn} the minimum CO core mass for carbon ignition is $\sim$ 1.05 \msun{}, whilst neon ignition requires ONe core masses $\sim$ 1.37 \msun{}. 

\subsubsection{Composition}

As the initial stellar metallicity decreases, stars attain higher central temperatures and luminosities during the main sequence to counteract fewer CNO seeds. This results in a larger He core mass for the same initial mass and also more massive cores during CHeB and hence resultant CO cores. Due to this the $M_{\rm{up}}$ values are seen to decrease with decreasing metallicity until reaching a plateau (or minimum e.g. \citealt{cassisi1993,bono2000}) at about Z=10$^{-3}$ to 10$^{-4}$. 
As seen in Figure~\ref{fig-mupmmass} the behaviour of $M_{\rm{mas}}$ with metallicity echoes that of $M_{\rm{up}}$ albeit with an offset of about $1.5\,$--$\,2.1$ \msun{} to higher initial masses.

Models that employ the strict Schwarzschild criterion for convective boundaries such as those from \cite{siess2007} typically produce the smallest HeB core and hence represent a reasonable upper limit to the values of $M_{\rm{up}}$ and $M_{\rm{mas}}$.
For near solar composition (Z=0.02) and using the strict Schwarzschild criterion the $M_{\rm{up}}$ and $M_{\rm{mas}}$ values are $\sim$ 9 and 11 \msun{} respectively  \citep{pap1,ritossa1999,siess2007,doherty2010,takahashi2013}. 
We note the other recent values of $M_{\rm{mas}}$ from the literature are in reasonable agreement with these values e.g. for $Z=0.015$ \cite{woosley2015} find 9 \msun{}, while \cite{jones2013} find 8.8--9.5 \msun{}. 

Even assuming the same convective approach during the pre-carbon burning phases, the $M_{\rm{up}}$ values vary considerably between studies. For example the work by \cite{gir00} find very low values of about 4.5--5\msun at Z=0.02. The causes of the differences are hard to attribute in some cases, as discussed in \cite{siess2007}.

The values of $M_{\rm{up}}$ and $M_{\rm{mas}}$ also show large variations due to the He content \cite[e.g.][]{bec79,bono2000}. Stars with larger initial He contents are more luminous and develop more massive convective cores during CHB. Their CHB lifetime is also substantially shorter due to both the reduced amount of H fuel and the hotter, larger cores which burn the fuel more efficiently. This larger core follows through to the CHeB phase resulting in a larger CO core which leads to a reductions of $\sim 1.6\,$--$\,2$ \msun{} in $M_{\rm{up}}$ and $M_{\rm{mas}}$ when enrichments of Y $\sim$ 0.15 are used \cite[e.g][]{bono2000,shingles2015}. 

\subsubsection{Overshooting}

Convective overshooting during the core H and He burning phases mixes in additional fuel to the core and increases the duration of these phases. It also increases the maximum size of the convective cores with this effect being more prominent during CHeB. Even moderate amounts of core overshoot e.g. $f_{\rm{over}}$ = 0.016 (see footnote~9) reduce the $M_{\rm{up}}$ and $M_{\rm{mas}}$ values by generally 2--2.5 \msun{} \cite[e.g][]{ber85,siess2007,gil07,poelarends2008,farmer2015}. This is highlighted in Figure~\ref{fig-mupmmass} by comparing the small/large open square value from \cite{siess2007} which are for models without/with overshoot respectively.

\subsubsection{Rotation}

Similar to the impact of overshooting, stellar rotation increases both the duration and the size of the convective core during CHB \cite[e.g][]{maeder2000,eks12}. This larger core is inherited during CHeB and hence we expect a larger CO core and presumably this would lead to a reduction in the initial mass for carbon ignition with increasing rotation rate. 
However, in their grid of intermediate mass Z=0.02 metallicity models with overshooting ($f_{\rm{over}}$) \cite{farmer2015} found that for a given initial mass, the CO core mass at carbon ignition was practically the same, irrespective of the initial rotation rate which ranged from $\Omega/\Omega_{\rm{crit}} = 0 \,$--$\, 0.5$ (their Figure~15). In this case it seems the rotation does not impact the $M_{\rm{up}}$ and $M_{\rm{mas}}$ values provided that overshoot is efficient enough.  

\subsubsection{Reaction rates}
\begin{figure}
 \begin{center}
 \includegraphics[width=8.5cm,angle=0]{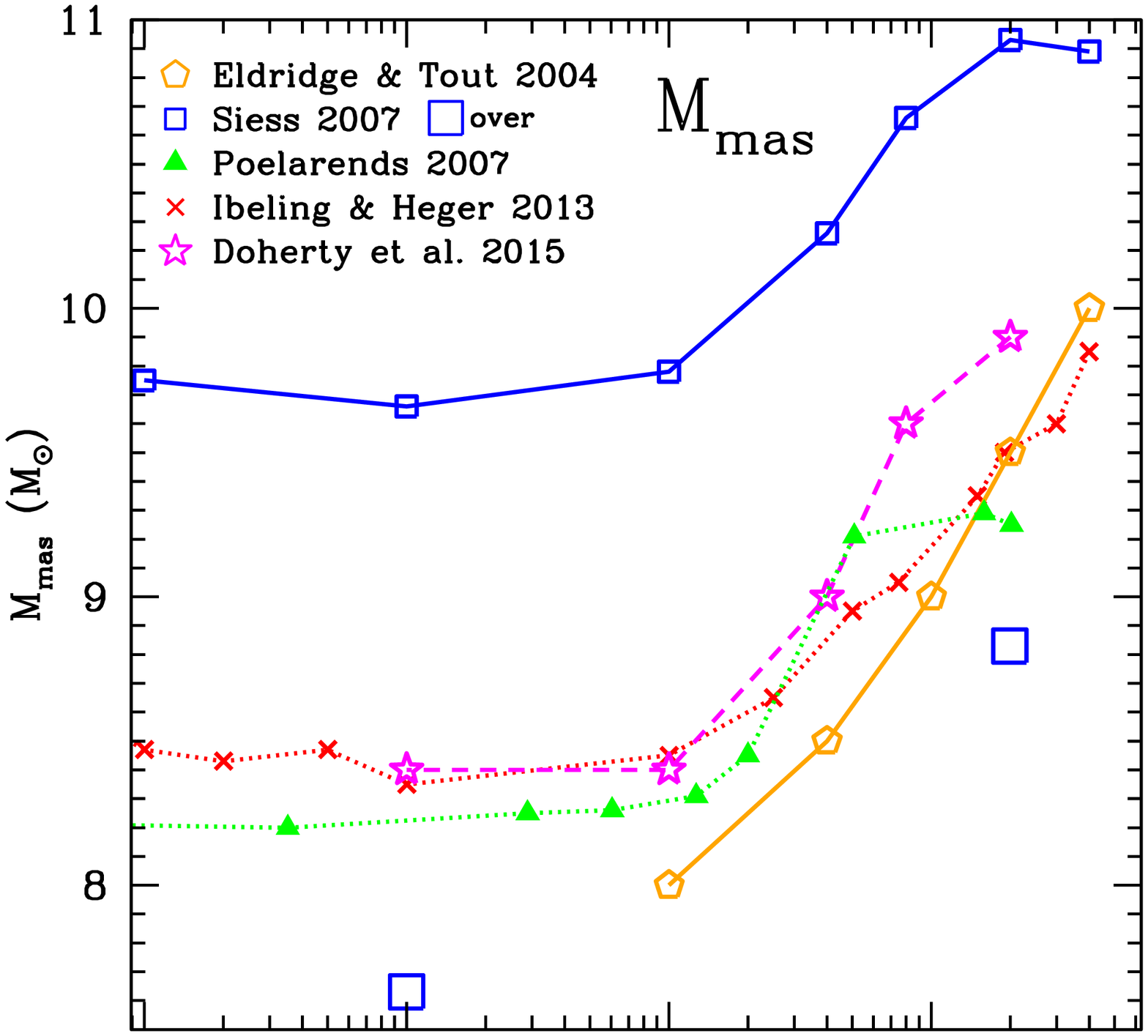}
 \includegraphics[width=8.5cm,angle=0]{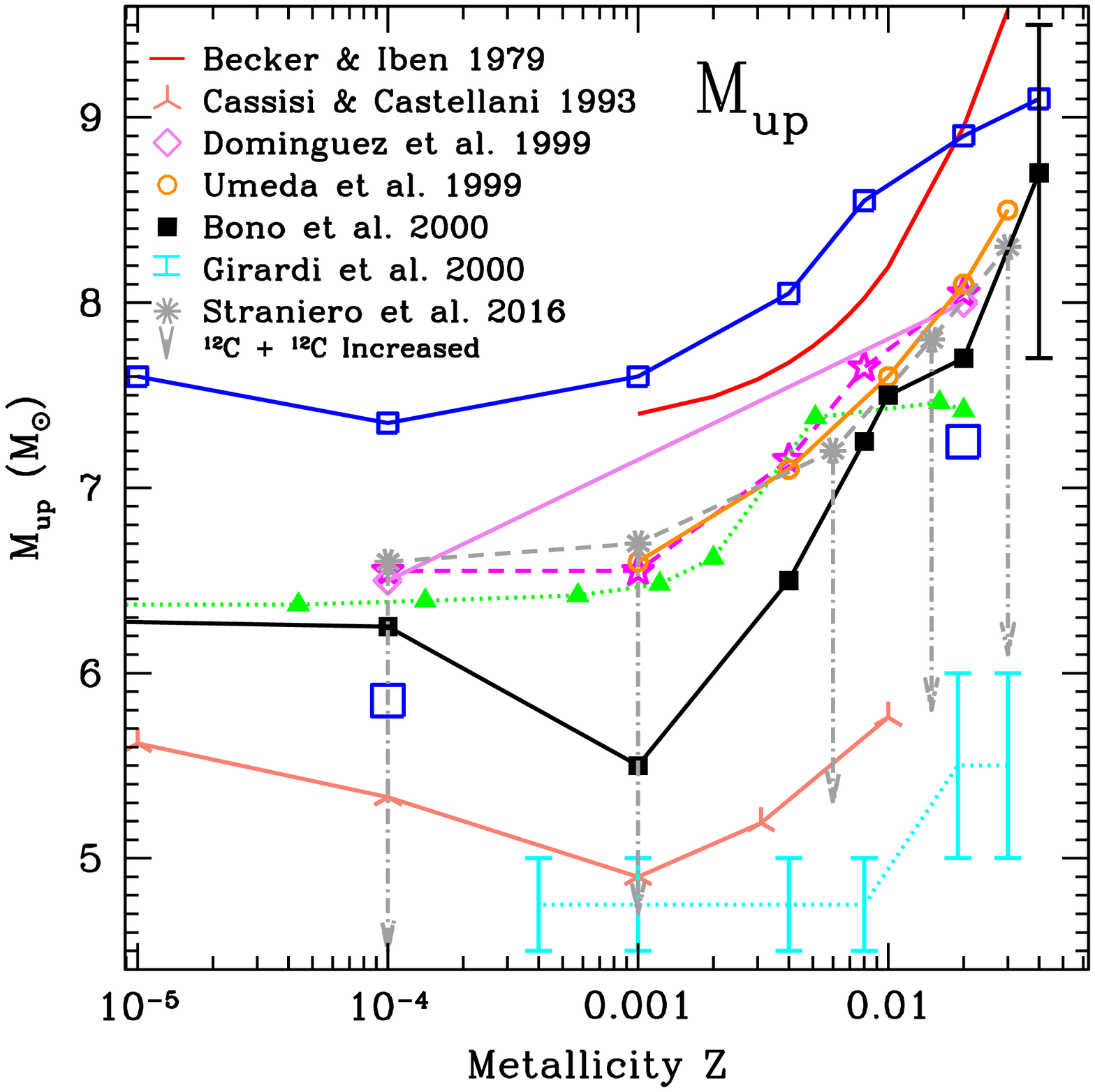}
 \caption{ Values for $M_{\rm{up}}$ (bottom panel) and $M_{\rm{mas}}$ (top panel) as a function of metallicity. Solid lines represent models calculated using the strict Schwarzschild condition for convective boundaries, dotted lines represent models calculated using mechanical overshooting during the core burning phases, whilst points joined with dashed lines represent models calculated using some other way of calculating the convective border, such as induced overshooting, a search for convective neutrality, or semiconvection. Values are from \cite{bec79}, \cite{bono2000}, \cite{cassisi1993}, \cite{doherty2015}, \cite{dom99}, \cite{eldridge2004}, \cite{gir00}, \cite{ibe13}, \cite{poe07}, \cite{siess2007}, \cite{straniero2016} and \cite{ume99}. The error bar on the Z=0.04 model from \cite{bono2000} represents the variation in $M_{\rm{up}}$ with initial helium content ranging from 0.29 to 0.37.} 
 \label{fig-mupmmass}
 \end{center}
 \end{figure}

The \chem{12}C + \chem{12}C  reaction rates can either hasten or delay the onset of C burning and hence alter the contraction time between He burning and C ignition. The CO core grows considerably during this phase and the impact of this should not be overlooked. The carbon burning \chem{12}C + \chem{12}C reaction rates can also modify the $M_{\rm{up}}$ value. It has been suggested that there exists a possible unknown/unmeasured resonance \citep{spillane2007} which would increase the reaction rate above that currently recommended (\citealt{cf88}, hereafter CF88) and lead to a reduction in the core mass which ignites carbon and hence reduce $M_{\rm{up}}$. \cite{straniero2016} examined the impact of including a narrow resonance at 1.45 MeV in the standard carbon burning rate. They found the minimum CO core mass for carbon burning was shifted from $\sim$ 1.06 \msun{} down to 0.95 \msun, and this resulted in a uniform decrease of $M_{\rm{up}}$ by about 2 \msun{} for their study over metallicities $Z= 0.0001\,$--$\,0.03$. This can be seen in Figure~\ref{fig-mupmmass} by the extent of the arrows representing models with the modified carbon burning rate. This reduction in $M_{\rm{up}}$ with increased carbon reaction rate is in agreement with the results of \cite{chen2014}. However with their maximum rate (1000 $\times$ CF88) they find a lesser decrease, at about 1.3 \msun{} (to $M_{\rm{up}}$ $\sim$ 5.3 \msun at metallicity Z=0.01).

In \cite{fra11} the $M_{\rm{mas}}$ value was seen to decrease by 1 \msun{} (to 7 \msun{} at metallicity Z=0.02) when the \chem{12}C + \chem{12}C reaction rates were enhanced by a factor of 10$^{5}$ compared to the standard CF88 rates. \\ 

Given the shape of the IMF and the decrease in $M_{\rm{mas}}$ with decreasing metallicity, we expect that the SN rate was higher in the past.
In summary, the two limiting masses $M_{\rm{up}}$ and $M_{\rm{mas}}$
are very uncertain and even with ``reasonable'' choices of input physics their values may vary by over 3 \msun. For example at close to solar metallicity (Z=0.02) $M_{\rm{up}}$ can vary between about to $5.5\,$--$\,9$ \msun.   
In Section~\ref{section5} we discuss the observational probes that are being used to aid in constraining these important mass boundaries.

 \subsection{Final fates of super-AGB stars - $M_{\rm{n}}$}\label{sub:ff}

 \begin{figure}
 \begin{center}
 \includegraphics[width=8.5cm,angle=0]{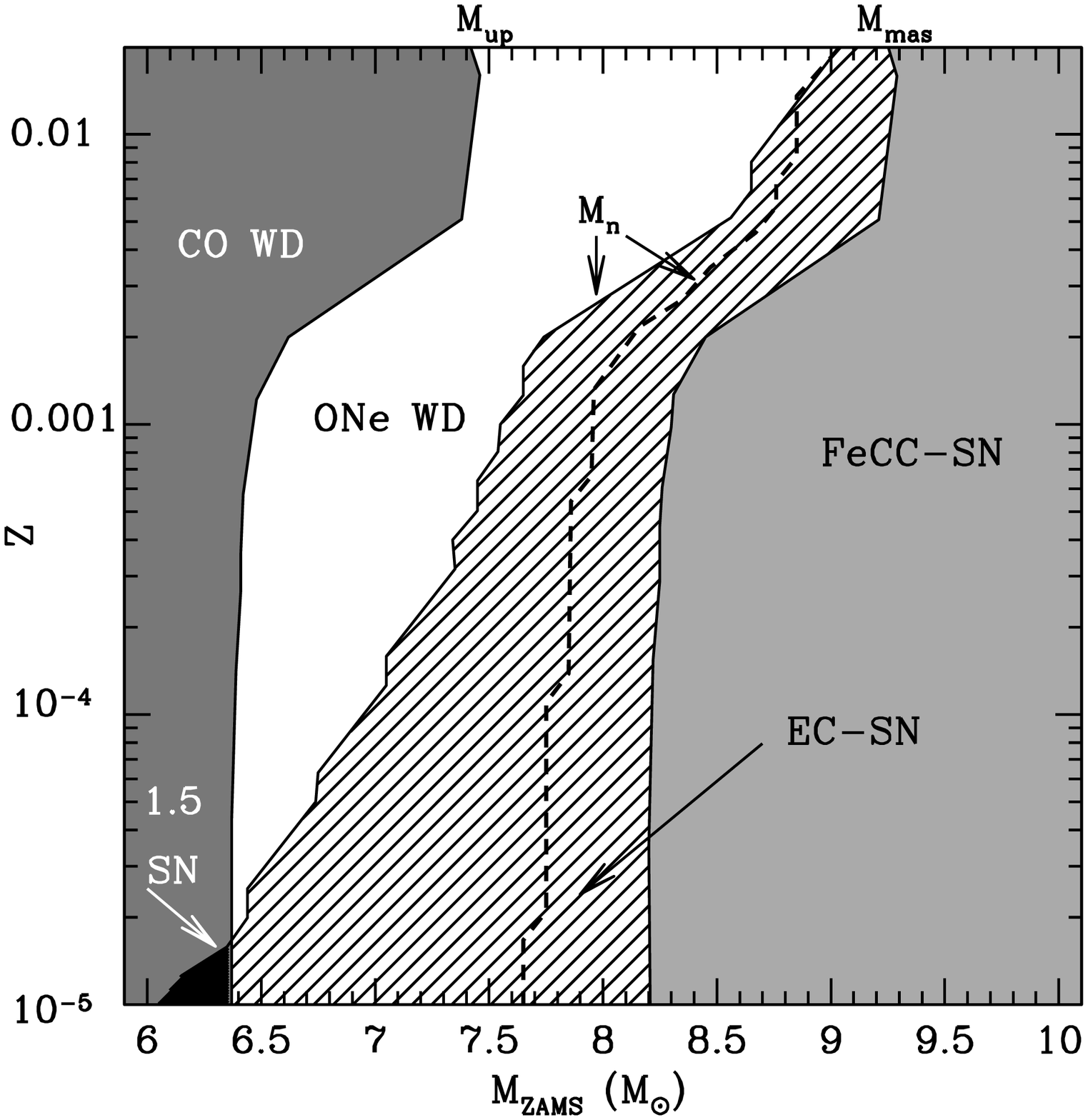}
 \includegraphics[width=8.5cm,angle=0]{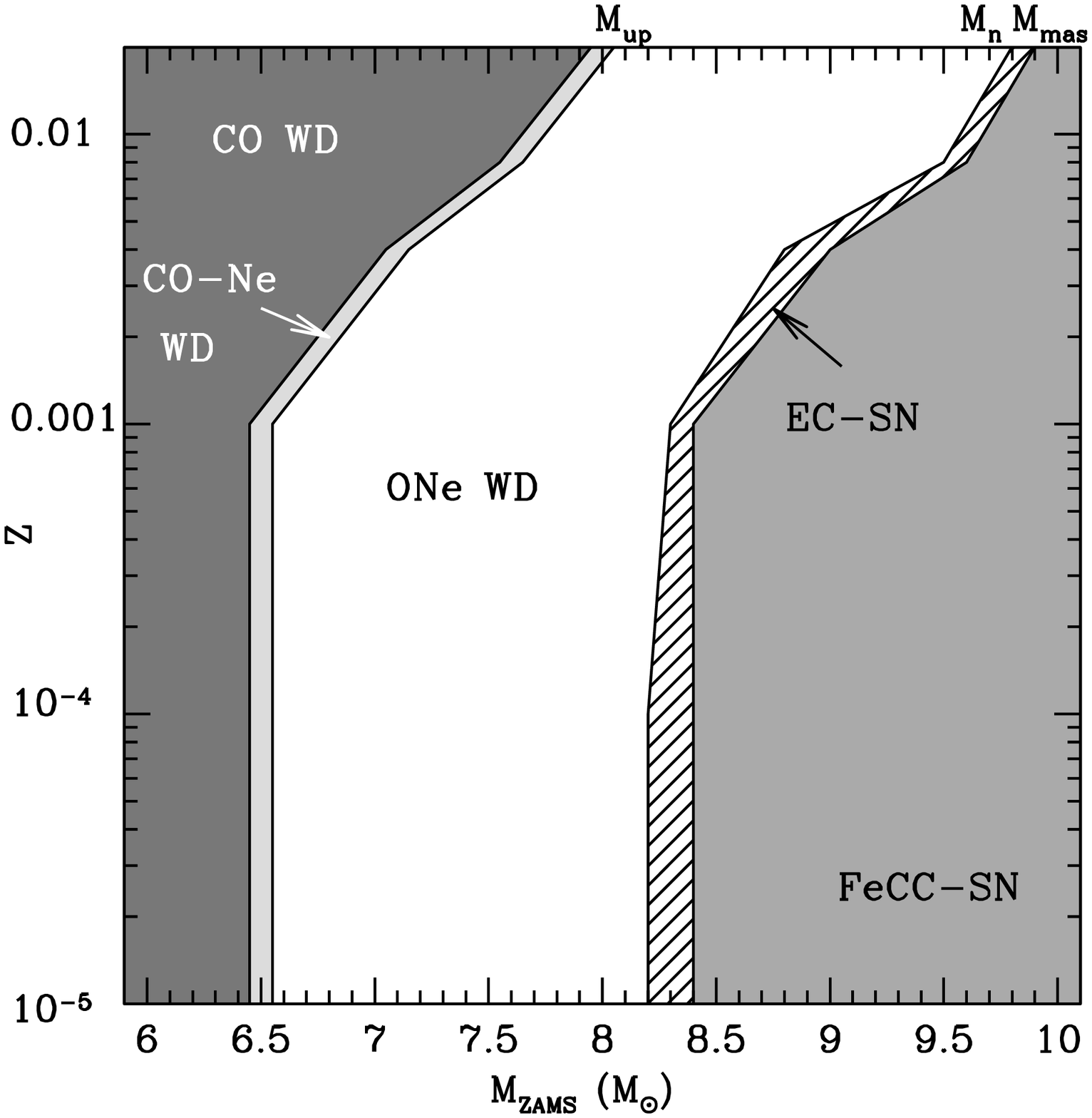}
 \caption{Final fates of intermediate-mass stars from \cite{poe07} (top panel) and \cite{doherty2015} (bottom panel). Solid lines delineate $M_{\rm{up}}$, $M_{\rm{n}}$ and $M_{\rm{mas}}$. The dashed line in the top panel represents the $M_{\rm{n}}$ value in the case where no metallicity factor is applied to the mass loss rate.  {The hatched region represents the width of the EC-SNe channel.} As mentioned in Section~\ref{critmass} there is a slight offset in the $M_{\rm{up}}$ and $M_{\rm{mas}}$ values between the two sets of models, with this due to the different method for treatment of convective boundaries during CHeB, with \cite{poe07} using convective overshooting via the method of \cite{her97} with an overshoot parameter of $f_{\rm{over}} = 0.016$, whilst in \cite{doherty2015} the search for convective neutrality approach of \cite{lat86} was used. We note that if no super-AGB stars become EC-SN then $M_{\rm{n}}$=$M_{\rm{mas}}$. The lowest metallicity examined in \cite{doherty2015} was Z=10$^{-4}$, but here we present new models for Z=10$^{-5}$ calculated using the same methodology as in the previous work.}
 \label{fig-ff}\end{center}
 \end{figure}

 After SDU or dredge-out has reduced the core mass to below $M_{\rm{Ch}}$, the final fate of super-AGB stars is dictated by the competition between core growth and 
 mass loss from the stellar envelope during the TP-SAGB phase. If the stellar wind removes the envelope prior to the core reaching $M_{\rm{Ch}}$ then the star will end its life as an ONe WD. Otherwise if the core growth is sufficient to reach $M_{\rm{Ch}}$ then the star will undergo an EC-SN and end its
 life as a neutron star. The boundary between these two differing final fates is called $M_{\rm{n}}$ the minimum mass for neutron star formation.\footnote{We note here that the end result of an EC-SN either as a neutron star or remnant ONeFe WD is debated e.g. \cite{Isern1991,canal1992,jones2016b}.}
 Here we describe the competing factors that determine the final fate including the complications from possible mass ejection events, and summarise the results from both synthetic/parametric and detailed  calculation that have examined this problem.

 The core growth rate is dictated by the outward movement of the H burning shell and progresses at $\sim$  10$^{-6}$ \msun yr$^{-1}$  \citep{ritossa1999,poelarends2008,siess2010,doherty2015} with typically faster growth rates in the more massive and/or metal-rich models.

 An important factor that influences the {\it {effective}} core growth 
 rate is the TDU.
 Unfortunately,  whilst the efficiency of TDU is a very important quantity, it is also one of the most poorly constrained aspects of AGB modeling, especially at larger core masses. It depends on many factors such as the resolution \citep{str97}, numerics \citep{sta04,sta06} and treatment of convective boundaries \citep{frost1996,her97,mow99,jones2016a}. 
 As mentioned in section~\ref{tpasgb}, in super-AGB stars the efficiency of TDU varies considerably between studies and ranges from $\lambda =0$ to $\lambda > 1$, typically being smaller for more massive models. There is some evidence for TDU in massive AGB stars of high metallicities. This is in the form of
 Rb over-abundances observed in bright O-rich AGB stars in the Galaxy and Magellanic Clouds \citep{gar06,gar09}. The existence of very luminous carbon rich stars in the Magellanic Clouds is suspected to arise from TDU events after the cessation of HBB \citep{fro98a,van99}. 
 Unfortunately we are yet to unambiguously identify any 
 super-AGB stars (see sections 5.1 and 5.3), and there are no constraints from lower metallicity objects, 
 because such stars have long since died. A standard tracer of AGB nucleosynthesis is \iso{99}Tc which is produced by neutron captures in the deep layers of the star. Technetium has no stable isotope, and its longest lived isotope is \iso{99}Tc with  a half-life of 0.21~Myr. Therefore the detection of Tc in the stellar spectra is the signature of recent nucleosynthetic activity and the presence of TDU.  
Unfortunately super-AGB stars are not expected to be produce large enough 
amounts of Tc for it to be observable (e.g. from the massive AGB stars study by \citealp{gar13}).

 Whilst the mass loss rate is fundamental to determining the final fates of super-AGB stars, unfortunately it is also highly uncertain especially at lower metallicities.  Mass loss in (super-)AGB stars is thought to be via pulsation aided dust-driven winds. Firstly large amplitude pulsations are responsible for forcing material to large enough radii and increasing the density enough for dust grains to form. The radiation pressure from the star is then able to accelerate these dust particles which also work to drag along the gas, with this leading to quite efficient mass loss \citep{wood79}.  Although there is no mass loss rate derived specifically for super-AGB stars it is common to use rates derived for lower mass AGB stars \cite[e.g.][]{vassiliadis1993,bloecker1995}, or for rates taken from red super-giant and O-rich AGB stars \citep{van05}. Using these prescriptions the mass loss rates in super-AGB stars are of the order 10$^{-4}$ to 10$^{-5}$ M$_\odot$ yr$^{-1}$. The mass loss rate for super-AGB stars at low metallicity is an unknown and we can only rely on and apply prescriptions which were derived using observations of solar metallicity, or moderately metal-poor stars. Due to their more compact structure, low metallicity stars are expected to have slower mass loss rates. \cite{kudri1987} proposed a metallicity scaling proportional to $\sqrt{Z/Z_\odot}$ in an to attempt to take this into account. We note however that this scaling was derived for radiative line-driven winds of hot luminous O stars whose conditions are quite unlike the cool super-AGB stars we are considering here.

 Another important factor which determines the mass loss rate in super-AGB stars is the envelope opacity.
 The use of low temperature molecular opacities that take into account the envelope composition variations during the super-AGB phase is crucial for cases where the envelope composition ratio C/O exceeds unity. The change in molecular chemistry when a star becomes carbon rich leads to an increase in opacity which results in a cooler and more extended stellar envelope and a higher mass loss rate \citep{mar02,cri07,ven10b,constantino2014,doherty2014b}. This effect may play an important role for low metallicity super-AGB stars. This is
 especially true for the most massive stars, with post SDU/dredge-out core masses closest to $M_{\rm{Ch}}$. These stars have often had their surface enriched in C due to dredge-out events and are carbon rich (with C/O $>$ 10 in some cases) already, 
 at the start of the TP-SAGB phase.
 
 The final fates of metal poor super-AGB stars also strongly influenced by the efficiency of SDU/dredge-out prior to the TP-SAGB phase. These processes are able to mix significant amounts of metals from the stellar interior leading to surface metallicities up to $Z=10^{-3}$, with the amount of enrichment increasing with stellar mass \citep{gilpons2013}. Furthermore, the nucleosynthesis and mixing processes which occur during the TP-SAGB also alter their total surface metallicity. As a consequence, their envelope opacity values, surface luminosities and radii become very similar to their higher Z counterparts. Envelope metallicity during the TP-SAGB is critical in terms of the strength of stellar winds, as higher Z objects are thought to be able to drive higher mass loss rates. 

 At low metallicities if the mass loss rate is sufficiently low 
 there is the possibility for stars with initial 
 masses $<$ $M_{\rm{up}}$ to be able to grow enough to reach $M_{\rm{Ch}}$ 
 and explode as Type 1.5 SNe \citep{ibe83,zij04,gil07,lau08,wood2011b}.

 In addition to standard super-wind mass loss from super-AGB stars there have also been two suggested potential mass expulsions events, from either the Fe-peak instability \citep{lau2012} or as a result of 
 global oscillations of shell-H ingestion (GOSH) events \citep{jones2016a}. These phenomena
 are expected to occur at different evolutionary phases, with the Fe-instability generally seen near the end stages of the TP-AGB when the envelope has reduced below about 3 \msun, whilst the GOSH could potentially occur prior to the start of the thermally pulsing phase during the dredge-out phase.
 These types of mass ejections make determining the final fates of super-AGB stars, in particular the most massive near the EC-SN boundary, quite problematic.

 Whilst the final fates of thermally pulsing super-AGB stars had been examined for individual models of $Z=0.02$ in the pioneering series of papers by Garcia-Berro, Iben and Ritossa \citep{pap1,pap2,pap3,pap4,ritossa1999} the global final fates problem for super-AGB stars was first tackled in a parametric fashion by \cite{poe07}, \cite{poelarends2008} and \cite{siess2007}, with  each of these studies using slightly different approaches.

 In \cite{poe07} a suite of synthetic models was computed exploring the rate of EC-SNe when using differing mass loss rates \citep{vassiliadis1993,van05}, efficiencies of TDU and a metallicity scaling included in the mass loss rate. The final fate results from their best estimate, which included the mass loss prescription from \cite{van05}, parameterised TDU from \cite{kar02} and metallicity scaling of \cite{kudri1987}, are shown in the top panel of Figure~\ref{fig-ff} (which is adapted from Figure 12 in \citealt{lan12}). This figure includes the critical mass limits delineating different evolutionary fates. We define $\Delta M_{\rm{EC-SN}}$ and $\Delta M_{\rm{ONe}}$ as the range of initial masses that produces  EC-SNe and ONe WDs respectively. For $Z=0.02$ this EC-SN channel is narrow with $\Delta M_{\rm{EC-SN}}$ $\sim$ 0.2 \msun, but at the lowest metallicity, all super-AGB stars would end life as EC-SNe, giving $\Delta M_{\rm{EC-SN}}$ $\sim$ 1.8 \msun, and leaving no ONe WDs. Interestingly at $Z=10^{-5}$ even the most massive CO cores are able to grow to $M_{\rm{Ch}}$ and explode as Type 1.5 SN. The cause of this increase in the EC-SN rate with decreasing metallicity is primarily the application of the metallicity scaling on the mass loss rate. In the case with no metallicity scaling applied to the mass loss (dashed line in Figure~\ref{fig-ff}) we find $\Delta M_{\rm{EC-SN}} \sim 0.25\,$--$\,0.55$ \msun with the wider range at lower metallicity. 
 
 Also critical to the width of the EC-SN channel is the occurrence of TDU, with lower efficiencies resulting in a wider channel. However, we expect the impact of the metallicity scaling upon the mass loss rate to be greater than the possible lack of TDU.
 In \cite{poelarends2008} the width of the EC-SN channel was explored for solar metallicity using a selection of commonly used mass loss rates \citep{reimers1975,vassiliadis1993,bloecker1995,van05} as well as the efficiency of TDU, finding $\Delta M_{\rm{EC-SN}}$ ranging from $\sim$ $0.2\,$--$\,1.4$ \msun. 
 For a rough estimation, the order of increasing mass loss rate is:  \cite{reimers1975}, \cite{van05}, \cite{vassiliadis1993} and  \cite{bloecker1995}, with approximate values for a typical metal-rich 
 super-AGB star being 0.2, 0.4, 0.8 and 3 $\times$ 10$^{-4}$ M$_\odot$ yr$^{-1}$ respectively (see Figure~7 in \citealt{doherty2014a}). 

  In \cite{siess2007} post SDU/dredge-out core masses were taken from detailed calculations and then the further evolution during the TP-SAGB phase was extrapolated based on the ratio of the average envelope mass loss rates $\dot M_{\rm{env}}$ to average effective core growth rates $\dot M_{\rm{core}}$ characterised by a $\zeta$ parameter defined as $\zeta=\mid{\frac{\dot M_{\rm{env}}}{\dot M_{\rm{core}}}}\mid$ .
   With best estimate values of $\dot M_{\rm{core}} = 5\times 10^{-7}$ \msun yr$^{-1}$, TDU efficiency $\lambda$ values between $0.3\,$--$\,0.9$ and $\dot M_{\rm{env}}$ = 5$\times$10$^{-5}$ \msun yr$^{-1}$ the $\zeta$ values range from $140\,$--$\,1000$. With these values the $\Delta M_{\rm{EC-SN}}$ ranged from $\sim$ 0.1 \msun{} ($\zeta$ $=$ 1000) to $\sim 0.25$--$0.6$ \msun ($\zeta$ $=$ 100) with these widths relatively constant over the entire metallicity range $Z= 0.04 \,$--$\, 10^{-5}$. When the metallicity scaling of \cite{kudri1987} was applied to the mass loss rate then a very similar result to that of \cite{poe07} was found with $\Delta M_{\rm{EC-SN}}$ increasing substantially at lower metallicity and at Z=10$^{-5}$ the ONe WD channel disappears entirely.  

 Using detailed evolutionary calculations \cite{doherty2015} confirmed the results of these earlier parametric studies. They found that when using reasonable mass loss rates \citep{vassiliadis1993}, without an explicit metallicity scaling but with efficient TDU, then the width of the EC-SN channel from TP-SAGB stars is narrow with $\Delta M_{\rm{EC-SN}}$ $\sim 0.1\,$--$\,0.2$ M$_\odot$ (bottom panel of Figure~\ref{fig-ff}). This Figure shows that the vast majority of stars that enter the thermally pulsing super-AGB phase will end life as ONe WDs, in sharp contrast with the favored set by \cite{poe07}.

 By taking the mass range of the EC-SN channel and weighting 
 it with an IMF, the importance and fractional contribution towards the overall core-collapse supernovae rate can be determined. 
 Using a Salpeter IMF and the EC-SN width and mass limits from the synthetic calculations of \cite{poe07}\footnote{We assume for our calculations that the maximum mass for a Type II SN is 18\msun{} based on the analysis of SN observations by \cite{smartt2015}} results in 5, 17 and 38 $\%$ of all Type II SN coming from the EC-SN channel for metallicities Z=0.02, 0.001 and 10$^{-5}$. 

 In contrast using the results from \cite{doherty2015} over these same 
 metallicities we find a far smaller percentage, of 
 about $2\,$--$\,5$\% of all Type II SNe will be EC-SNe. 
 This large variation in frequency of EC-SNe highlights the importance of constraining the mass loss rate at low metallicity.

 We note here that models that undergo core overshooting will have far 
 smaller envelopes (by as much as $2\,$--$\,3$ \msun) to remove 
 in order for a star to avoid the EC-SN channel. This will result in fewer SNe for models with larger values of overshooting. 

 \subsection{The binary channels towards EC-SNe}\label{sec:binary}

 The calculations presented in the previous section have shown that 
 the EC-SN channel for single super-AGB stars strongly depends on the mass loss 
 rate and the efficiency of the TDU, both of which are highly uncertain and
 remain poorly constrained. Using standard prescriptions for the wind 
 mass loss which has no explicit dependence on metallicity\footnote{The  metallicity is {\it implicitly\/} included via its effect on the structural variables that appear in the formula.},
 \cite{siess2007}, \cite{poelarends2008} and \cite{doherty2015} showed that the initial mass 
 range for single stars to evolve toward EC-SNe is narrow, being
 of the order of $0.1\,$--$\,0.5$\msun. However, this picture does not 
 consider binary evolution which opens new channels for the formation 
 of such SNe and ONe WDs. This is particularly relevant considering 
 the high fraction of stars having a companion \cite[]{Raghavan2010} 
 and that about 70\% of all stars more massive than 15\msun{} will 
 interact at some point with a binary companion \cite[]{Sana2012}.

 We will start our discussion with the accretion onto an ONe WD in a 
 short period system. These systems represent the massive counterparts of 
 cataclysmic variables and follow a similar evolutionary scenario for their 
 formation but with different initial conditions \cite[for a review on the  
 formation of cataclysmic variables see e.g.][]{ritter2012}. The evolution starts with a super-AGB 
 progenitor and a lower mass companion. If the initial period is long 
 enough, mass transfer by Roche lobe overflow starts when an ONe core 
 is formed and the super-AGB star has entered the TP-SAGB phase. 
 Because the TP-SAGB star has a deep convective envelope, mass transfer is 
 likely to become dynamically unstable\footnote{Note however that if the mass 
 ratio is less than $\sim 1.2-1.5$ \cite[]{webbink1988}, mass transfer by 
  Roche lobe overflow  from a convective star does not become dynamically unstable. Indeed, 
 in such circumstances, the rapid reversal of the mass ratio can stabilize 
 the system because mass transfer now occurs onto a more massive companion 
 leading to an increase in the separation.}  resulting in the formation of a 
 common envelope. When the mass transfer timescale becomes shorter than the 
 thermal timescale of the accreting component, the gainer star cannot 
 assimilate the incoming material and the matter soon engulfs the binary 
 system. The friction of the stars with the gas will produce a
 spiralling-in with transfer of orbital angular momentum and potential 
 energy to the envelope. The outcome of this process is either the merger of 
 the two stars if the envelope is not ejected sufficiently rapidly or the 
 formation of a short(er) period binary system if the common envelope is 
 rapidly dispersed. This phase is short lived ($\sim 1000$yr) and during that 
 short period of time, the companion is not expected to accrete a significant 
 amount of mass. Today, the determination of final orbital parameters is 
 still subject to large uncertainties associated with our understanding of 
 the energetics involved \cite[for a review of common envelope evolution see][]{ivanova2013}.

If the system avoids merging, we are left with a detached system.
The hot core of the super-AGB star ionizes 
the expanding envelope, enabling it to shine for $\sim 10^4$yr 
as a planetary nebula. Subsequently the 
companion may fill its Roche lobe driving a second phase of 
(reversed) mass transfer. This can be initiated either by 
the star expanding as a result of its nuclear evolution, 
or by the loss of orbital angular momentum due to gravitational wave  
emission and/or magnetic coupling.

On the other hand, for shorter periods, mass transfer may be initiated 
while the super-AGB star progenitor is on the red giant branch.  
Because of the extended convective envelope, this case B  Roche lobe overflow  is unstable and, 
for the reason mentioned before, a common envelope develops. 
The H-rich envelope of the primary is ejected and the outcome is the formation of 
a short period system composed of a naked He star and 
a low mass companion. The evolution of this post-common-envelope binary has been investigated 
by \cite{law1983} and detailed stellar models were 
computed by \cite{dominguez1993} and \cite{GPGB2001,GPGB2002}. These simulations 
indicate that a second episode of mass transfer is triggered 
after the onset of He shell burning (referred to as case BB) and because 
the He star has a radiative envelope, the mass loss from the primary is 
stable.
In their study of a 10\msun{} primary with a lower mass companion, \cite{GPGB2001} showed that the second mass transfer episode is  
stable and that the system detaches when carbon ignites leading to the 
formation of a cataclysmic 
variable with an ONe WD. On the other hand, starting with a 9\msun{} initial 
model, the primary looses so much mass that it does not ignite carbon and ends up as a CO WD \citep{GPGB2002}.

The fate of the accreting WD depends mainly on the mass accretion rate
 \cite[e.g.][]{nomoto1982}: if it is too low, recurrent H shell 
 flashes (nova outbursts) eject more mass than has been accreted and the 
 accretor loses mass. On the other hand if the accretion rate is higher than the core 
 growth rate (which is controlled by the H-burning shell), the accreted 
 envelope expands and a common-envelope may ensue
 followed by a spiral-in phase. In the intermediate regime, H burning is 
 steady and stable accretion allows the ONe core to grow. When the central 
 density reaches $4\times 10^9$g\,cm$^{-3}$ for an ONe core mass of $\sim 
 1.37$\msun{} \cite[]{nomoto1984}, electron capture reactions on \chem{24}Mg, 
 \chem{24}Na and \chem{20}Ne are successively  activated. They induce the collapse of the 
 white dwarf and the heat released by $\gamma$-ray emission ignites oxygen 
 burning. 
 The outcome of this accretion-induced collapse  depends on whether or 
 not the timescale for electron capture (which induces contraction) is 
 shorter than the timescale associated with the nuclear energy 
 release by oxygen burning (which produces expansion). 

 The competition 
 between these two processes is sensitive to the density 
 where nuclear burning is ignited, which 
 in turn depends on the adopted input physics \cite[e.g.][]{Isern1994} 
 and initial conditions (in particular the initial WD mass). If the density 
 is too low, the propagation of the burning front can lead to complete 
 disruption of the core and not to a collapse. \cite{nomoto1991}, \cite{Isern1991} 
 and more recently \cite{Schwab2015} showed that the collapse of an ONe 
 core leads to the formation of a neutron star rather than a thermonuclear 
 explosion because in these WDs the density at the time of oxygen ignition is 
 high enough for electron captures to proceed faster than the other 
 nuclear reactions contributing to the nuclear energy production.  
 The  2D hydrodynamical simulations 
 by \cite{Dessart2006,Dessart2007}, \cite{Janka2008} 
 and \cite{wanajo2011} confirm the results of 
 the 1D models, in that oxygen deflagration does not 
 lead to a thermonuclear explosion. This means that 
 the outcome of the accretion induced collapse of ONe WDs 
 is most likely a neutron star.
However, the recent 3D hydrodynamical simulations by \cite{jones2016b} suggest that, depending on the efficiency of semiconvective mixing, these cores may not collapse to form neutron stars but instead undergo a thermonuclear explosion which results in a bound ONeFe WD remnant.

 Binary evolution opens new possibilities and in particular that 
 the formation of an ONe WD or EC-SN does not necessarily require 
 a super-AGB star progenitor. The proposed scenario considers the merging 
 of two CO WDs. The formation of such systems has been investigated by 
 \cite{Iben1984} and \cite{Webbink1984} and involves one or two common envelope 
 episodes. Once the two WDs are formed in a short period system, 
 angular momentum loss by gravitational wave radiation brings the 
 two cores closer to each other. Eventually, the lower mass WD, 
 which has the largest radius, starts overfilling its Roche lobe and is 
 eventually disrupted by the strong tidal forces. The material distributes in 
 a thick accretion disk around the massive WD and mass transfer of hot CO 
 material onto the surviving companion begins. The mass transfer is then 
 expected to be very high, of the order of $10^{-5}$\myr, close to the 
 Eddington limit. As initially investigated by \cite{Saio1985}, in this 
 double degenerate scenario the fast accretion leads to off-center 
 ignition of carbon and the inward propagation of a burning front that 
 incinerates the CO WD into an ONe core that may eventually collapse 
 into a neutron star if it is massive enough. However, this picture is 
 oversimplified. Hydrodynamical simulations 
 \cite[e.g.][]{Guerrero2004,Pakmor2012} reveal that soon after the disruption 
 of the WD, a ``quasi-static'' configuration develops 
 in which the cold WD core is enshrouded in a hot, rapidly 
 rotating CO envelope surrounded by a thick Keplerian disk. 
 Using these new and more realistic initial conditions, \cite{Yoon2007} 
 re-investigated the conditions for off-centre carbon ignition and showed that it 
 depends on the temperature in the hot envelope, the
 mass accretion rate and the mass of the WD. 
 Recent 3D SPH simulations \cite[e.g.][and reference therein]{Sato2016} also 
 predict that the outcome, accretion-induced collapse or Type Ia explosion, 
 depends sensitively on the initial mass ratio.

 The evolution of single super-AGB stars tell us that the SDU  
 can significantly reduce the mass of the H-exhausted core below the 
 Chandrasekhar limit as illustrated in Figure~\ref{fig:SDUp}. Models also show 
 that the core growth during the subsequent TP-SAGB phase is very modest and 
 that very few stars reach the conditions for EC-SNe. 
 This prompted \cite{podsi2004} to suggest 
 that binary interaction may be able to 
 remove the envelope before the SDU occurs and thus allow more 
 super-AGB stars to become EC-SNe.

 In some early papers, \cite{nomoto1984,nomoto1987} investigated the fate of 
 He cores representative of the evolution of 8-10\,\msun{} models and showed that 
 for He core masses in the range 2-2.5\msun, the evolution proceeds toward 
 EC-SNe. Therefore a new channel to EC-SNe will open if, at the end of a case A or B  
 mass transfer, the super-AGB donor keeps a helium core mass in the range 
 2-2.5\,\msun. However, \cite{podsi2004} also pointed out that this simple 
 picture does not take into account the effects of binary interactions which 
 can substantially alter the stellar structure. In particular, the size of 
 the helium core depends on the mass of the hydrogen envelope. Previous 
 studies \cite[e.g.][]{wellstein1999,gilpons2003} indeed show that the helium 
 core mass can be dramatically reduced in short period systems compared to 
 the evolution of a single star. Furthermore, binary interactions contribute 
 to redistributing the angular momentum inside the star. This is likely to 
 generate additional mixing which will also affect the He core mass.
 This promising scenario remains to be explored with self-consistent models, 
 so that the range of initial periods and stellar masses can be identified. 

 Investigating the origin of unusual fast and faint optical transients, 
 \cite{tauris2013,tauris2015} studied the evolution of binary systems 
 composed of a neutron star orbiting a helium-star companion. This 
 initial set-up is the result of previous binary evolution which can be 
 summarized as follows. The starting point is  two main sequence stars with 
 the primary being a typical B star 
 ($3 M_\odot \lesssim M_1 \lesssim 20 M_\odot$).
 Depending on the initial period and mass ratio, the more massive component 
 undergoes  Roche lobe overflow  while on the main sequence (case A) or after core 
 H-exhaustion (case B). As a result of this conservative mass transfer, the 
 system is composed of a naked He-star (the initially more massive star 
 stripped of its H-rich envelope) and a relatively massive ($8 M_\odot 
 \lesssim M_2 \lesssim 20 M_\odot$) main sequence companion (which has 
 accreted a substantial fraction of the primary's envelope) in a wide orbit. 
 The more evolved He star eventually explodes as a supernovae (possibly an 
 EC-SN!) leaving a neutron star remnant. If the explosion does not disrupt 
 the system, the neutron star can accrete some of the wind from the secondary  
 (likely a Be star) and may show up as an X-ray source.
 Thereafter the secondary fills its Roche lobe but because of the extreme 
 mass ratio, mass transfer is dynamically unstable and leads to common-envelope 
 evolution. After spiral-in, and provided merging is avoided, the final system 
 consists of a short period neutron star orbiting the 
 naked He core of the secondary. 
 The subsequent evolution is similar to the previous one. After core 
 He exhaustion, the He-star expands and a new episode of mass transfer begins 
 \cite[]{habets1986,dewi2002} which is temporarily halted during central C 
 burning. At this stage, the exchange of mass is highly 
 non-conservative because the mass transfer rate is 3--4 orders of magnitude 
 larger 
 than the Eddington accretion limit of the neutron star (a few $10^{-8}$\myr). 
 In their study, \cite{tauris2015} showed that by gradually increasing the 
 initial period (from 0.06 to 2.0 days) and the mass of the He-star, the 
 remnant of the He-star varies from CO WD to ONe WD, 
 to neutron stars formed by EC-SN or 
 FeCC-SN as a consequence of mass transfer occurring during core 
 helium burning (case BA), He shell burning (case BB) or beyond (case BC). 
 In some cases, the mass transfer runs away and a common-envelope episode follows 
 leading to a merger or the formation of a very tight system. In this 
 scenario, EC-SN will mainly be observed as weak Type Ic SNe.

 \section{Super-AGB Star Nucleosynthesis}\label{sec:ns}

 A variety of mixing episodes can occur in intermediate mass 
 stars prior to the TP-(S)AGB phase. 
 Super-AGB stars of high and moderate metallicities undergo the first dredge-up event that mixes to the surface material from regions that have undergone partial hydrogen burning in which the CNO cycle is active but with only marginal activation of the heavier hydrogen burning cycles/chains. There are 
 increases in the surface abundances of \chem{4}He, \chem{14}N, \chem{13}C, \chem{17}O  and to a lesser extent  increases in \chem{23}Na, \chem{21}Ne, and \chem{26}Mg. Simultaneously there is a decrease in the surface abundances of H, \chem{7}Li, \chem{12}C, \chem{15}N and \chem{18}O.

 As with the FDU, SDU mixes to the surface  species involved in H burning, 
 with the main nucleosynthetic signature being a very large enhancement 
 of \chem{4}He by up to $\sim$0.1 in mass fraction and also significant surface 
 enrichment of \iso{14}N and \iso{23}Na.
 In addition to the enrichment from standard SDU, models with corrosive SDU also 
 increase the surface abundance of \iso{12}C, and in some cases \iso{16,18}O and \iso{22}Ne. 
 Dredge-out events also have the ability to enrich the surface in products of partial He burning, in particular \iso{12}C. As mentioned in \cite{doherty2015} and \cite{jones2016a} the \iso{13}C produced via proton capture on \iso{12}C will subsequently undergo the \iso{13}C($\alpha$,n)\iso{16}O reaction with the neutrons produced leading to potentially significant heavy element production.

 Once the star reaches the TP-super-AGB phase the nucleosynthesis is dictated by hot bottom burning and (potentially) third dredge-up.

  \subsection{Hot bottom burning}

\begin{figure}
  \begin{center}
  \includegraphics[width=5.5cm,angle=0]{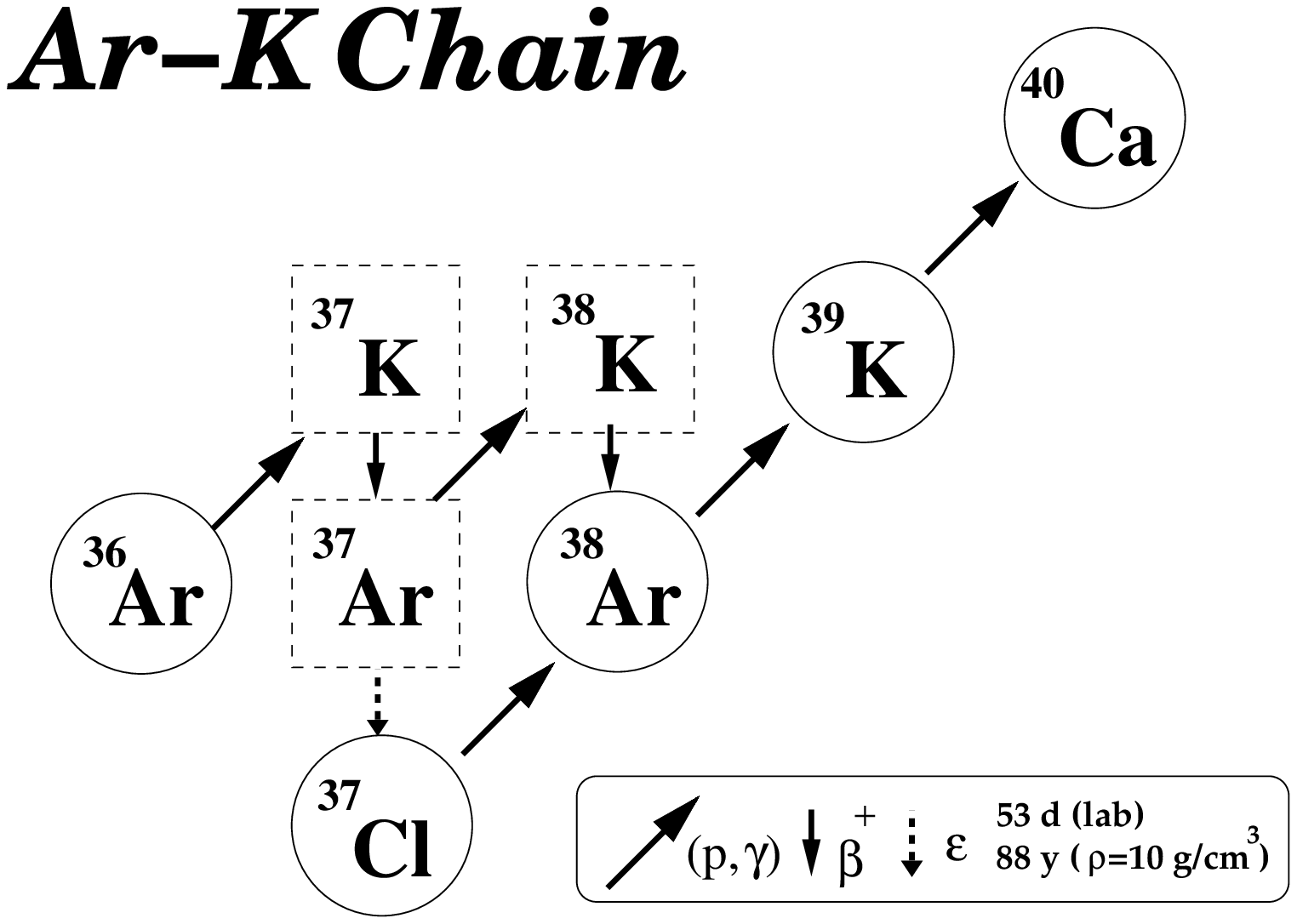}
  \includegraphics[width=8.5cm,angle=0]{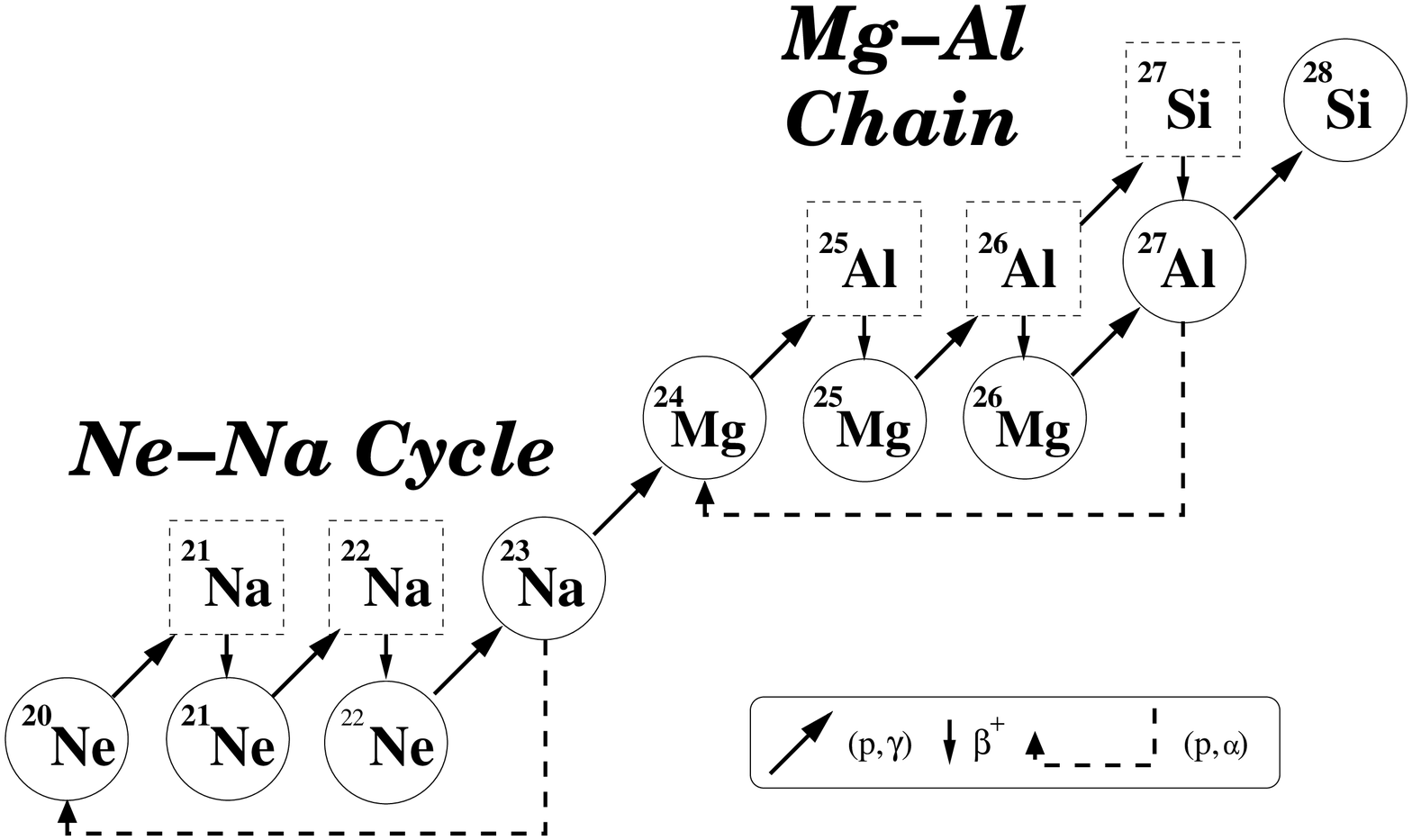}
  \includegraphics[width=7.5cm,angle=0]{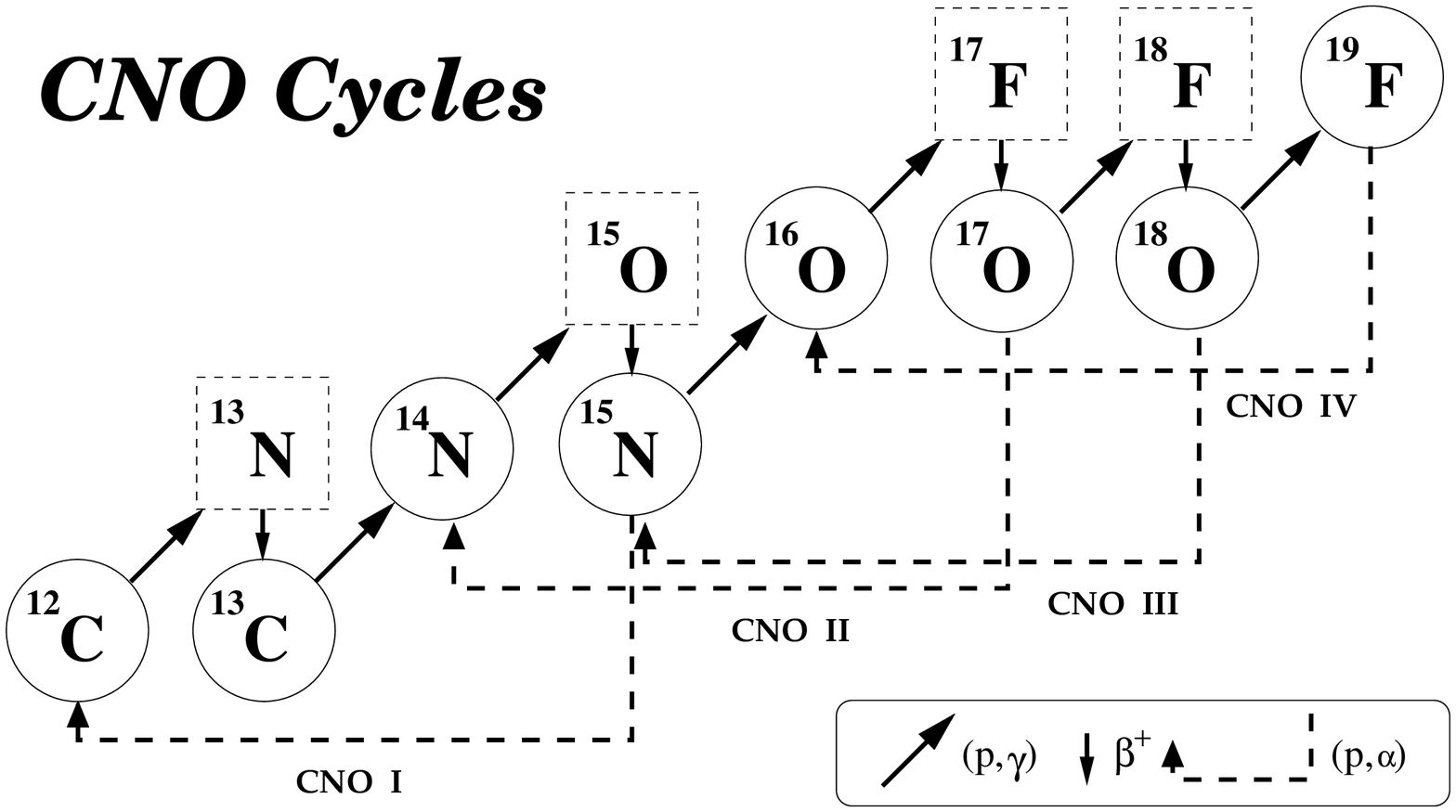}
  \caption{A section of the chart of the nuclides 
  (with atomic mass number on the x-axis and proton number 
  on the y-axis) showing the CNO and Ne-Na cycles and Mg-Al and Ar-K chains.}
  \label{fig-hbb}
  \end{center}
  \end{figure}

  Hot bottom burning (hereafter HBB) occurs during the interpulse phase 
  when the material in a very thin region at the base of the convective 
  envelope is hot enough to undergo nuclear burning. An equivalent, 
  and perhaps more intuitive, way to think of this
  is that the bottom of the convective envelope extends 
  into the top of the H-burning shell.
  The maximum temperature found at the bottom of the 
  envelope is a function of initial metallicity and mass, with more 
  massive and/or metal-poor models achieving higher temperatures. 
  For super-AGB stars this temperature ranges from about $100\,$--$\,160\,$MK with a density of about 10 g cm$^{-3}$. With such high temperatures there is activation of the CNO, Ne-Na, Mg-Al chains/cycles and potentially 
  proton-capture reactions involving heavier species such as Ar and K (see Figure~\ref{fig-hbb}). 

  Another consequence of HBB in super-AGB stars is the
  activation of the Cameron-Fowler mechanism \citep{cam71}. 
  Here \chem{7}Be is created from \chem{4}He(\chem{3}He,$\gamma$)\chem{7}Be 
  and is quickly mixed to a cooler region where it undergoes  
  electron capture to form \iso{7}Li.
 Super-AGB stars can indeed become very lithium rich, with A(\chem{7}Li) 
  up to $\sim$ 4.5\footnote{A(\chem{7}Li) = log (n[Li]/n[H]) + 12, 
  where n is number abundance}, for a short time at the start, or just prior to the TP-(S)AGB phase. But once the \iso{3}He in the envelope is depleted the Li production ceases and Li is efficiently destroyed.
 HBB also produces  \iso{4}He, but unless the TP-SAGB phase is very extended, the main contribution to the surface of \iso{4}He will remain SDU/dredge-out.

 Activation of the CNO cycles leads to the production of \chem{13}C and \chem{14}N to the detriment of \iso{12}C, with this efficient destruction being able to decrease the C/O ratio to below unity in the cases where the surface has been enriched from corrosive SDU, dredge-out or efficient TDU. However, at high temperatures ($>$ 95MK) the \iso{16}O(p,$\gamma$)\iso{17}F channel opens and depletes \iso{16}O. In the most metal-poor/massive super-AGB star models this can lead to the creation of a carbon star (C/O $>$ 1) not from C enrichment but from the depletion of O \citep{siess2010}. 
 The \iso{12}C/\iso{13}C and \iso{14}N/\iso{15}N number ratios reach their equilibrium values.

 Within the Ne-Na cycle very hot HBB leads to the destruction of \chem{21,22}Ne and \chem{23}Na to the benefit of the already very abundant \chem{20}Ne. 

 For super-AGB stars the most obvious result from the Mg-Al burning is the very large reduction of \chem{24}Mg which is almost completely destroyed to form \chem{25}Mg. \chem{26}Al is also produced \citep{siess2008,doherty2014a} but at these high temperatures the \chem{26}Al(p,$\gamma$)\chem{27}Si($\beta^{+}$)\chem{27}Al channel opens which bypasses the \chem{26}Mg. 
  As the temperatures increase even further (T $>$ 110 MK) there is activation of the \chem{27}Al(p,$\gamma$)\chem{28}Si reaction \citep{ventura2011b}.

  If the HBB temperatures in super-AGB stars are extreme ($>$ 150 MK) the argon-potassium (Ar-K) chain (refer Figure~\ref{fig-hbb}) may also be activated \citep{ventura2012}, with this chain of particular relevance to the globular cluster abundance anomaly problem (refer to Section.~\ref{sub:GC}). 

 We note here that the temperature at the base of the convective envelope is strongly influenced by the theory of convection that is used. Models such as those by \cite{ventura2005} and \cite{ventura2011b}, which use the full spectrum of turbulence (FST) approach \citep{can91}, attain higher temperatures for the same initial mass when compared to models which use standard mixing length theory.

 The nuclear reaction rates are also a source of great uncertainty in these heavier proton chains and this can lead to substantial differences between results depending on which reaction rates are used, in particular the linking reactions between the chains/cycles such as \iso{23}Na(p,$\alpha$)\iso{20}Ne. 

 Near the end of evolution when the envelope mass drops below about $1\,$--$\,2$ \msun, the temperature reduces below the critical value required to sustain HBB and this process ceases.

 \subsection{Third dredge up}\label{sub:TDUns}

 Within the intershell convective zone during a thermal pulse 
 the main product is \chem{12}C produced by the triple-$\alpha$ 
 reaction with subsequent production also of the $\alpha$ capture species \chem{16}O, \chem{20}Ne, \chem{24}Mg, and \chem{28}Si. The abundant \chem{14}N in the intershell from the preceding CNO cycling is converted to \chem{22}Ne through the reactions \chem{14}N($\alpha,\gamma$)\chem{18}F($\beta^+\nu$)\chem{18}O($\alpha,\gamma$)\chem{22}Ne. Due to the large temperature in the helium burning intershell convective zone, the \chem{22}Ne($\alpha$,n)\chem{25}Mg and \chem{22}Ne($\alpha,\gamma$)\chem{26}Mg reactions are activated. There is a substantial neutron flux resulting from the \chem{22}Ne($\alpha$,n)\chem{25}Mg reaction with neutron densities of up to 10$^{14-15}$\;n/cm$^{3}$. These free neutrons can be captured by Fe seeds within the convective pulse and lead to heavy element production via the slow (s) neutron capture process.  
 
 Relevant to heavy element production in super-AGB stars is the 
 occurrence of hot TDU  events, where the temperature at the base of 
 the convective envelope during the TDU is still high enough for 
 nuclear burning to be active \cite[e.g][]{chieffi2001,her04a}. 
 From theoretical predictions \citep{goriely2004} these hot TDUs 
 are expected to inhibit the formation of \iso{13}C pockets. 
 With no \iso{13}C pocket formation, in the standard picture, 
 the main neutron source in S-AGB stars for the 
 production of elements heavier than Fe 
 is assumed to be the \chem{22}Ne neutron 
 source during the convective thermal pulses. Recently \cite{jones2016a} 
 found proton ingestion episodes during the later thermal pulses 
 creating \chem{13}C which may lead to $i$-process nucleosynthesis \citep{cowan1977} via the \chem{13}C($\alpha$,n)\chem{16}O neutron source.  We shall discuss the heavy element production yields from super-AGB stars in Section.~\ref{subsec:heavy}. Owing to the very narrow mass of the convective He shell instability, 
 even with a substantial number of thermal pulses, the total amount of material dredged to the surface of super-AGB stars is quite modest, at most about 0.1 \msun.

 \subsection{Yields}\label{sec:yields}

 \subsubsection{Light Elements}

 Super-AGB star yields for the production of elements lighter than Fe 
 have only quite recently become available. There is now a variety of calculations from different research groups which cover a large range of masses and metallicities including a wide variety of input physics. 
 Grids of nucleosynthesis calculations of super-AGB stars have been produced from three main groups: those using \textsc{starevol} \citep{siess2010}, \textsc{aton} \citep{ventura2013,dicriscienzo2016} and \textsc{monstar/monsoon} \citep{doherty2014a,doherty2014b}. For full details of choices of stellar model parameters the reader should refer to these publications. Here we briefly describe the major differences between the different set of models.

 In contrast to the model grids using \textsc{monstar}, those using \textsc{starevol} and \textsc{aton} do not undergo TDU events and therefore show the abundance patterns of pure HBB. Another significant difference between models is that \textsc{starevol} and \textsc{monstar} use standard mixing-length theory with an alpha value calibrated to solar value. The \textsc{aton} models  utilize the FST \citep{can91} model for convection, which produces
 higher temperatures during HBB, with the result that 
 these models undergoing more advanced nucleosynthesis. 

The mass loss prescriptions also vary.  The standard model sets for \textsc{starevol} and \textsc{monstar} both use the mass loss prescription from \cite{vassiliadis1993}, whilst \textsc{aton} use the more rapid \cite{bloecker1995} rate with an $\eta$ value of 0.01.
 Slower mass loss rates lead to more TDU enrichment (if TDU is occurring) and the longer duration on the super-AGB gives HBB more time to process material. This, together with higher temperatures at the bottom of the convective envelope, will lead to more advanced nucleosynthesis. Another important factor is the nuclear reaction rates, especially for the heavier proton capture reactions.

 A common feature of all super-AGB models at all metallicities is the large production of HBB products \iso{13}C and \iso{17}O as well as \iso{4}He  with the bulk amount coming from efficient SDU. Lithium is temperamental and is either destroyed or produced in super-AGB stars dependent primarily on the mass loss rate. When HBB begins we see production of Li, but once the \iso{3}He is all used, the destruction of Li dominates. So if the mass-loss is sufficiently high that the star ends its life early, before the \iso{3}He is all destroyed, then the star may be a nett producer of Li. If the mass-loss rate is lower, and the Li is destroyed before most of the mass is ejected, then the Li yield is negative. \citep{ven10a,doherty2014a}. \iso{14}N is increased at all dredge-up events and is also greatly increased through HBB.  The TDU products \iso{12}C, \iso{16}O and \iso{22}Ne are also subsequently processed via HBB. 

 Models of very (and extremely) metal-poor super-AGB stars create similar isotopes to their metal rich counterparts. However, in lower metallicity models that experience TDU, such as those by \cite{doherty2014b}, one
 find positive yields of the species \iso{12}C, \iso{16}O, \iso{15}N, \iso{28}Si, which is not the case for the more metal-rich super-AGB stars.

\begin{figure}
 \begin{center}
 \includegraphics[width=8.5cm,angle=0]{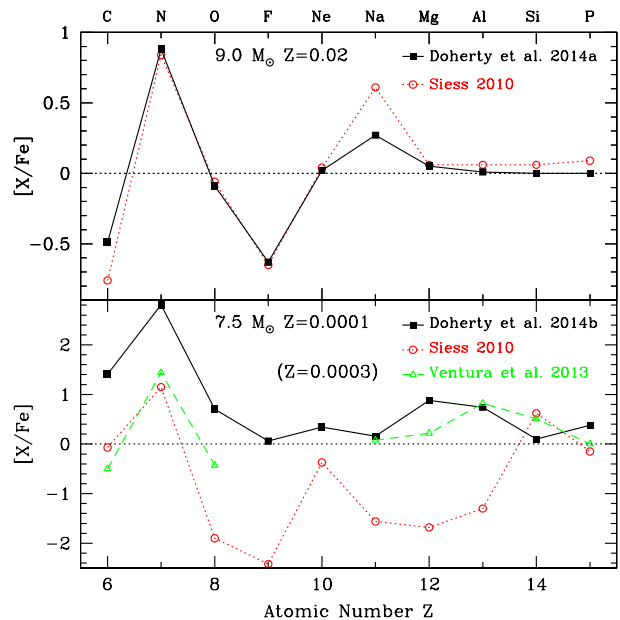}
 \caption{Comparison of a selection of light element yields for models of 9.0 M$_\odot$ $Z=0.02$ and 7.5 M$_\odot$ $Z=0.0001$ (or $Z=0.003$) from \protect{\cite{doherty2014a,doherty2014b}}, \protect{\cite{siess2010}} and \protect{\cite{ventura2013}} Note the change of scale for the y-axis between panels.}
 \label{fig-comparison}\end{center}
 \end{figure}

 In Figure~\ref{fig-comparison} we compare light element super-AGB star yields (in [X/Fe])\footnote{where [A/B]$ =\log_{10}(n$(A)$/n$(B))$_*-\log_{10}(n$(A)/$n$(B))$_\odot$} for models with $Z=0.0001$ (bottom panel) and close to solar composition (top panel) in common between the studies by \cite{doherty2014a,doherty2014b} and \cite{siess2010}. We also include the slightly higher metallicity ($Z=0.0003$) model from \cite{ventura2013}.  Clearly seen is the close agreement found between studies for metal-rich super-AGB star yield predictions. This is due to nucleosynthesis in metal rich super-AGB stars being primarily driven by HBB.  However, as the metallicity decreases the differences between results from different groups begins to increase. This is due to a variety of factors, such as variations in the HBB temperatures and choice of nuclear reaction rates but is primarily related to the occurrence or not of TDU, with the relative TDU contribution higher at lower metallicity. At the very low metallicity the yield of many major elements such as C, O, F, Ne, Na and Mg differ so much between calculations that there is no consensus on whether these elements are either produced or destroyed within super-AGB stars.

 We stress that although agreement between results from different groups at high metallicity is somewhat comforting, the principal test of the validity of our results will come when we can compare against observations, which requires us to positively identify super-AGB stars.

For details of nucleosynthesis in extremely metal-poor (Z $\le 10^{-5}$) and primordial super-AGB stars we  refer to the review by Gil-pons (2017) in this volume.

 \subsubsection{Heavy elements}\label{subsec:heavy}

\begin{figure}
 \begin{center}
 \includegraphics[width=8.5cm]{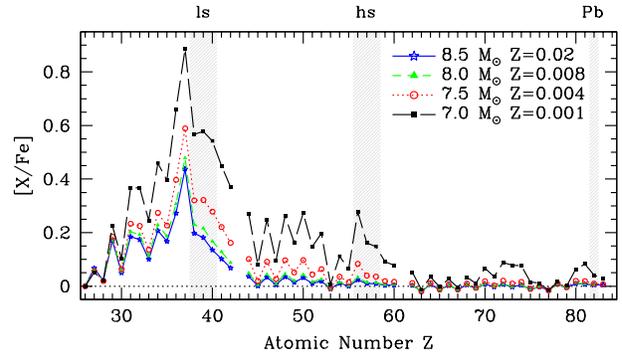}
 \caption{Heavy element nucleosynthesis yields for super-AGB stars for a range of metallicities (in [X/Fe]) all scaled to the solar abundances of \cite{asplund2009}. The breaks in the distribution are for the elements Tc (Z=43) and Pm (Z=61) which have no stable isotopes. The shaded regions represent the elements used to represent the three s-process peaks $ls$, $hs$ and Pb. The maximum production is for the element Rb (Z=37).}\label{fig-sprocess}
 \end{center}
 \end{figure}

 Due to both the numerical complexities and time consuming nature of the calculations, currently published heavy element nucleosynthesis yield predictions of super-AGB stars are limited to a small selection of individual models in \cite{fishlock2014b}, \cite{shingles2015} and \cite{karakas2016}\footnote{Note also the super-AGB star nucleosynthesis calculations for selected species in \cite{karakas2012} and \cite{lugaro2014}}. In these works heavy element production is limited to predominantly Rb and in not large quantities.  This nucleosynthesis is illustrated by Figure~\ref{fig-sprocess} in which we present a new set of super-AGB heavy element (s-process nucleosynthesis) yields for a range of metallicities $Z=0.02-0.001$. These calculations are based on the evolutionary models from \cite{doherty2014a,doherty2014b} and use the 324 species nucleosynthetic network detailed in \cite{lugaro2014}. These models are chosen as representative of intermediate-mass super-AGB stars and all have very similar core masses ($\sim$1.15 \msun), thermally pulsing durations ($\sim 6 \times 10^4$ yrs) and number of thermal pulses ($\approx$ 100). 
 As mentioned previously, the high temperature within the helium burning shell during the convective thermal pulses leads to strong activation of the \chem{22}Ne neutron source with very high peak neutron densities N$_n$ $\sim 10^{14}$\,cm$^{-3}$ reached. However, with the very short thermal pulse duration, the integrated neutron exposure $\tau$ is not very high, at most $\sim$ 0.04 mbarn$^{-1}$. This results in a large production of Rb (Z=37), Kr (Z=36) and light s-process elements Sr, Y, and Zr (Z=38, 39, 40), and a small synthesis of heavy s-process elements and Pb. The heavy element abundance patterns are similar between models of the different metallicities even with the associated large variation in availability of Fe-seeds.  This is due to the low neutron exposure which inhibits formation of substantial amounts of elements past the light s-process peak as well as the small overlap factor between successive thermal pulses which leads to little build up of heavy elements to be subsequently reprocessed.
 Even with their large number of thermal pulses, heavy element production within super-AGB stars (at least of the moderate metallicities presented here), is reasonably modest, with the yield of no heavier than Fe element exceeding unity in [X/Fe].
 . This is due primarily to the small mass contained within each convective thermal pulse ($\sim$ 5$\times$ 10$^{-4}$ M$_\odot$) and therefore considerable dilution of s-process enriched intershell material within a massive envelope.  

The super-AGB models of \cite{jones2016a} found proton ingestion during TPs where protons from the envelope were making contact with the convective HeB during the thermal pulse. This leads to rapid production of \iso{13}C, which then undergoes the \iso{13}C($\alpha$,n)\iso{16}O reaction.  The neutrons produced during this event are available to then form heavy elements via the intermediate capture process.  \cite{jones2016a} estimated the potential i-process heavy element production within super-AGB stars to be of the order of 1--2 dex, comparable to the maximum 1--2.5 dex production from the \iso{22}Ne source as estimated by  \cite{doherty2014a,doherty2014b}.

 \subsection{Contribution to chemical evolution}

 Due to a lack of stellar yield calculations, super-AGB stars had long 
 been missing in galactic chemical evolution studies, with 
 this mass range either treated
 by interpolating between lower mass AGB stars and massive 
 stars \cite[e.g][]{rom10a,kob11}, or even neglected entirely.
 Now, with the recent availability of grids of super-AGB stellar yields 
 (from a variety of research groups and utilising quite different input physics), we
 are beginning to be able to answer the question of how important are these stars, within a galactic perspective.

\begin{figure}
 \begin{center}
 \includegraphics[width=8.5cm,angle=0]{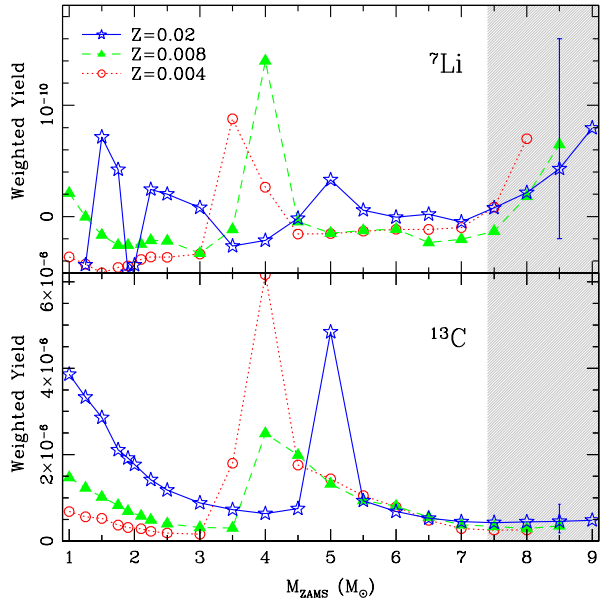}
 \caption{Stellar yields weighted by the \cite{kro93} IMF, with the shaded regions representing the mass range for super-AGB stars. For masses lower than 6 \msun{} AGB yields are from \cite{karakas2010}. The error bars on the 8.5 M$_\odot$ Z=0.02 are mass loss tests cases from \cite{doherty2014a}.}
 \label{fig:weighted}
 \end{center}
 \end{figure}

 In  \cite{doherty2014a} metal-rich super-AGB yields were weighted by a standard IMF and compared to the contribution from lower mass AGB stars 
 to assess their relative importance. 
 This showed that whilst metal-rich super-AGB stars are large producers of isotopes such as  \chem{4}He, \chem{13}C, \chem{14}N, \chem{17}O, \chem{22}Ne and \chem{23}Na, from a galactic context their contribution is minimal.  In Figure~\ref{fig:weighted} we have selected two illustrative isotopes, \chem{7}Li and \chem{13}C. For \chem{13}C the overall contribution is practically negligible compared to that from the intermediate mass AGB stars.  However, depending on the choice of mass loss rate, super-AGB stars at high/moderate metallicities may make a contribution to 
 the Galactic inventory of \chem{7}Li, \chem{25,26}Mg and \chem{27}Al.
 Super (and massive) AGB stars also contribute a non-negligible amount ($\approx$ 10 per cent) to the galactic value of the radioactive isotope \chem{26}Al \citep{siess2008,doherty2014a}.
 As mentioned in Section~\ref{sec:yields}, the yields of super-AGB stars vary quite considerably between the results from different research groups. However this is less pronounced at higher metallicities. This makes these models reasonably robust. 
 
 Based on current model predictions, super-AGB stars (at least of mid to high metallicity) are not expected to make a substantial contribution to the heavy elements in the Galaxy. This is of course dependent on the choice of mass loss rate, with a larger contribution expected for a slower mass loss rate.  In super-AGB stars there is considerable dilution of s-process enriched intershell material within the massive envelope. Yet super-AGB stars may still make an important contribution to the light s-process elements, in particular Rb, Sr and Y in the early Galaxy. Currently the impact of the possible heavy element (i-process) nucleosynthesis in dredge-out events \citep{doherty2015,jones2016a} and proton ingestion during thermal pulses \citep{jones2016a} awaits detailed nucleosynthesis calculations. 

 \subsection{Globular cluster abundance anomalies}\label{sub:GC}

 The ``abundance anomaly problem'' in globular clusters  involves explaining the origin of the  unusual compositions present in a substantial fraction (often a majority) of the stars in globular clusters. These patterns
 are not seen in field stars, and include, for example, He enrichment and anti-correlations in the
 element pairs C-N, O-Na, Mg-Al and Mg-Si \citep{car09a,car09b,bra10a}. 
 These abundance patterns are characteristic of the results of hot hydrogen burning \citep{den90,pra07}. Crucially in the majority of clusters there 
 is no variation, from star to star, in the  Fe and heavy (s-process) element 
 abundances \cite[e.g.][]{yon06b,yon08a}.
 There also seems to be 
 a near constancy in the total C+N+O abundance \cite[e.g.][]{smi96,iva99}, 
 with some exceptions \cite[e.g. NGC1851,][]{yon09}.  
This would limit the amount of dredged-up material as this contains primary carbon (and its burning products) whereas 
 H burning simply cycles CNO elements among each other, leaving C+N+O constant.
 
 The leading theory to explain these anomalies is that globular clusters are made of multiple generations of stars, with the anomalous stars being formed from the enriched material from a first generation of stars (for a review refer to \citealt{gra12}). Many candidates sites have been proposed as the source of the enriching gas such as: (super-)AGB stars \citep{cot81,ventura2001,der08}; rapidly rotating massive stars \citep{dec07b}; massive binary stars \citep{dem09}; super-massive stars \citep{denissenkov2014} and novae \citep{smith1996,mac12}. 

 Currently each of the suspected polluters has problems matching certain observed abundance patterns \citep{gra12} as well as larger problems such as
 the mass budget for the different populations and the polluted material \cite[e.g.][]{bastian2015}. The entire picture of multiple stellar generations seems incompatible with observations from young massive clusters (the best current day globular cluster analogues) that show a lack of available gas, and no evidence for multiple bursts of star formation \cite[see][and references therein]{bastian2015c}.

 Whilst there are problems with the (super-)AGB scenario, as well argued by \cite{renzini2015} and \cite{dantona2016}, using solely nucleosynthetic considerations super-AGB and massive AGBs are perhaps the only barely plausible progenitor candidate remaining. This is based on their ability to (qualitatively) reproduce many of the necessary features, such as high helium content (primarily from SDU) and Li production.  They also reach high enough temperatures to activate not only the CNO cycles but also the Ne-Na and Mg-Al chains/cycles.
 
 Observations of Mg-K anti-correlations in the low metallicity globular clusters NCG2419 and NCG2808 \citep{Mucciarelli2012,cohen2012,mucciarelli2015} have been explained as being produced from (very) hot hydrogen burning and the argon-potassium (Ar-K) chain \citep{ventura2012}.  This is shown in  Figure~\ref{fig-hbb}. Under laboratory conditions the half life of \iso{37}Ar to electron capture is 53 days whilst at conditions found in super-AGB star envelopes with $\rho$ $\sim$ 10 g/cm this half-life is increased to 88 years. 
 Due to this, \cite{iliadis2016} argued that the main path to \iso{39}K would be via the \iso{36}Ar(p,$\gamma$)\iso{37}K($\beta^+\nu$)\iso{37}Ar(p,$\gamma$)\iso{38}K($\beta^+\nu$)\iso{38}Ar(p,$\gamma$)\iso{39}K channel. \cite{iliadis2016} used Monto Carlo nuclear reaction network calculations 
 to identify a temperature and density regime that could explain all the abundance patterns in the cluster NGC2419, in particular the Mg-K anti-correlation. They ruled out the majority of polluter candidates but concluded that super-AGB stars may be a viable solution if the HBB temperatures could be increased slightly,\footnote{Nova also may reach the correct temperature/density conditions, however without any detailed nova model calculations at low metallicity this needs further exploration and remains quite speculative.}  such as results from the use of the FST theory of convection.

 Probably the major nucleosynthetic problem with the (super-)AGB scenario currently is the fine balance of conditions required to produce  \iso{23}Na instead of destroy it. 
 To achieve the high depletion of O (and Mg) that is observed, the duration of the TP-(S)AGB phase needs to be quite extended. However this leads also to destruction of \iso{23}Na, whereas the stars depleted in O are enhanced in \iso{23}Na.  A remedy to this problem could come from a decrease in the reaction rate of \iso{23}Na(p,$\alpha$)\iso{20}Ne destruction channel by a factor of 2--5 \citep{ven06}. We note however that this value is outside the current uncertainties. An experimental re-evaluation of this reaction rate is sorely needed.

 Another nucleosynthetic obstacle for the (super-)AGB star scenario (and all other polluter candidates) is the production of too much \iso{4}He to match the observed spread in the majority of clusters \citep{bastian2015b}.  Helium in massive AGB and super-AGB stars is produced primarily from SDU, with a smaller contribution from HBB.  These processes are quite independent, 
 so there is some potential scope to modify the \iso{4}He production. 

 Another important and often overlooked point is the impact of rotation on the surface composition. \cite{dec09} showed that rotation in intermediate mass stars may act like deep (corrosive) SDU and increase the surface in carbon (and total metallicity), which would make (super-)AGB stars unviable polluter candidates.

 As we saw in Section~\ref{sec:yields} the predictions of 
 super-AGB star nucleosynthesis at low metallicity (applicable 
 to the majority 
 of globular clusters) vary widely between different 
 research groups. For super-AGB star 
 models to match as the polluter we must fit rather 
 strict evolutionary  constraints. These are: no, 
 or very inefficient TDU;
 very advanced HBB, driven either by more efficient convective mixing (e.g. 
 through use of FST), or standard mixing-length theory with increased mixing 
 length $\alpha$; a relatively slow mass loss rate. 
 In addition, as discussed in \cite{pumo08}, the most massive 
 super-AGB stars, which undergo dredge-out events, must all end 
 their lives as EC-SNe to avoid polluting 
 the cluster with C enriched material.

 However,  if (super-)AGB stars are indeed the source of the material from which the later generation(s) formed this would prove invaluable to constraining theoretical modelling of this class of star.

 \subsection{Dust production}\label{sub-dust}

 AGB and super-AGB stars make substantial contributions to the galactic dust inventory \cite[e.g.][]{gehrz1989,Matsuura2009,sch14}. 
 The type of dust formed during the (S-)AGB phase is determined by the surface composition of the star, primarily the C/O ratio, and is classified into two main groups: either carbon-rich or oxygen-rich dust.  As seen in the previous sections, due to HBB the surface composition in super-AGB stars is typically C/O $<$ 1 for the majority of the TP-SAGB phase. Thus super-AGB stars are expected to produce silicates (olivine, pyroxene $\&$ quartz), alumina dust and no or very little carbon dust \cite[e.g.][]{ventura2012a}. 
 Dust yields have now been calculated for super-AGB stars over the wide range of metallicities $Z=0.001-0.018$ \citep{ventura2012a,ventura2012b,ventura2014,dell2017}.

 The amount of dust generated depends critically on the density in the stellar outflow, with more rapid mass loss leading to higher density and greater dust production. This results in dust mass yields (and silicate dust grain sizes) which are correlated with initial mass, with the most massive super-AGB stars producing $\sim$ 10$^{-3}$ to 10$^{-2}$ \msun  with grain size  $\lesssim$ 0.15 $\mu$m \citep{dell2017}.
 Models of O-rich super-AGB stars from \cite{ventura2012b} find the amount of dust production is strongly metallicity dependent, with decreasing dust yields at lower metallicity. This is due to the reduced availability of the key elements Si, Al, Fe and O at lower metallicity, as well as the far lower O abundance due to the more efficient HBB at low metallicity. Since these elements are not produced in substantial amounts within the stars themselves, dust formation is expected to be inhibited. At the metallicity of $Z=3\times 10^{-4}$ \cite{dicriscienzo2013} found the Si abundance was too low for non-negligible silicate production. They suggested a metallicity threshold of $Z=0.001$ as the limit for silicate production for super-AGB stars.
 As yet, there are no dust yield calculations from models that have undergone dredge-out and become C-rich at the start of the super-AGB phase, nor of super-AGB star models which have become carbon rich due to either 3DU or very efficient O destruction. These models would result in substantially different dust chemistry and warrant exploration.


 \section{Discussion and Conclusions}\label{section5}

 Here we discuss some of the observational probes and studies that can and are being used to aid in the understanding of the evolution of stars that bridge the low/high mass divide.  

\subsection{Observational constraints}

 As mentioned in the introduction, a fundamental problem 
 is that there are at present no definite detections of super-AGB stars.
Super-AGB stars with their very rapid mass loss rates are expected to be dust enshrouded OH/IR stars and more luminous than their lower mass AGB star counterparts. However, models that undergo more efficient convective mixing (e.g. 
 through using the FST theory of convection or 
 a larger mixing-length within the MLT formulation) will attain higher luminosities, making clear identification between a massive AGB and a super-AGB star practically impossible. Furthermore, from a nucleosynthetic point of view a low mass super-AGB star and a massive AGB star are expected to produce very similar element distributions, in particular for the heavier than Fe elements.   
At present abundance determinations in massive (potentially super-)AGB stars are limited to a small sample of stars in the Galaxy and Magellanic clouds \citep{gar06,gar07,gar09}. These stars are found to be Rb-rich, which is suggestive that they have undergone TDU events. Determining the exact amount of enrichment is problematic as it has been shown that accounting for an extended  circumstellar envelope can substantially modify the abundance determinations in this class of star \citep{zamora2014}. 
Super-AGB stars may also be identified in the large infrared surveys of Magellanic clouds\cite[e.g.][]{kra17}, or in young stellar clusters.

Super-AGB stars also share the same stellar luminosities and surface temperatures as the slightly more massive red super-giants.  Hence a careful analysis of red super-giant surveys may yield super-AGB star candidates. One such object is the star HV2112 \citep{lev14}, which has been suggested potentially as a super-AGB star, or even  a Thorne-\.Zytkow object \citep{tout2014}. Their stellar variability may be a key to distinguish between 
red super-giants and super-AGB stars with super-giants 
found to have lower amplitude pulsations, with maximum variations 
of about 0.5 mag \citep{woo83,Groenewegen2009}.

 \subsubsection{Low mass channel}

 Planetary nebula abundances may be used as probes of the resultant nucleosynthesis in super-AGB stars \cite[e.g.][]{ventura2015,ventura2016b,gar16}. However as these stars represent only a relatively scarce (and short-lived) mass range we expect only a small number of such PNe.

 White dwarfs can be used to explore the evolutionary properties of super-AGB stars in a variety of ways, such as WD mass distributions/population studies, initial-final mass relations  and analysis of individual massive WDs. 

We note that observations of ultra-massive\footnote{Massive WDs have masses greater than 0.8 \msun{} whilst ultra massive have masses greater than 1.1\msun} white dwarfs are particularly difficult due to their relative scarcity, inherently low luminosity due to more compact nature, as well as more rapid cooling \citep{alt07}.
Recent works by \cite{cum16a,cum16b} and \cite{raddi2016} have derived semi-empirical initial-final mass relations for ultra-massive WDs using observations from young open clusters. However, the procedure to derive an initial-final mass relation is quite involved and is affected by quite a range of uncertainties. Firstly, the WD effective temperature and gravity are determined from observations and the mass is then determined from a mass to radius relation. Next the cluster age is found from isochrone fitting with this procedure reliant on stellar evolution models with their inherent uncertainties. The initial mass of the WD is determined by estimating the cooling time, which is dependent on composition e.g. CO \citep{sal97,sal10} or ONe \citep{alt07}, and then subtracting this from the cluster age to find the stellar lifetime. Lastly this duration is compared to theoretical stellar models to derive the initial mass. With the stellar lifetimes of super-AGB stars quite short ($\sim 20\,$--$\,60$ Myr) and dependent on rotational mixing and overshoot among other effects, the determination of an initial mass for a ultra-massive WD will be hampered by large uncertainties. Hence this method seems unlikely to be able to accurately determine the exact upper mass limit for WD formation. 
Based on current estimates of the initial-final mass relation the maximum mass for a WD  progenitor is estimated to be in range $\sim$ 7.5 to 10 \msun{} or more  \citep{sal09,wil09,cum16a}. 

 Theoretical calculations from single or binary intermediate mass stars predict a veritable zoo of massive WDs including: CO, CONe, ONe, Si and  ONeFe WDs. Yet observationally distinguishing these cores may prove to be practically impossible. Two ultra-massive ($\sim$ 1.1 \msun{}) O-rich \citep{gan10} WDs have been observed.\footnote{The first WD with an atmosphere of O, Ne and Mg has been recently discovered \citep{kepler2016} . However, unexpectedly the surface gravity of this WD corresponds to a mass of $\approx$ 0.56 \msun, far below that expected from theoretical predictions. One possible explanation is that this WD formed via a binary channel.}
  In principle differentiating between the CO or ONe core composition of a massive WD can be accomplished from asteroseismology of pulsating WDs \citep{cor04}. Another way to probe WD interiors comes from either classical or neon novae which are hosted on CO and ONe WDs\citep{jos98,wan99,gil03,dow13} respectively, however these observations will most likely probe only the outer core.

 \subsubsection{High mass channel}

There is the 
potential to learn about the final fates of intermediate mass stars through studies of SNe and their remnants. 
The most common approach is to identify particular events and make direct comparison between models and observed properties such as chemistry, mass-loss history, light curves etc. The overall SN rate may be used to estimate the minimum mass for a SN. In addition, it may also be possible to use predicted nucleosynthetic properties to determine the frequency of certain types of SN events.

Have we already seen the death throes of a supernova from a super-AGB star? 
With their H rich envelopes, single super-AGBs are expected to explode as Type II SN. EC-SNe are expected to have lower explosion 
energies of $\approx$10$^{50}$ erg and very small Ni masses $\sim$ 1--3 $\times$ 10$^{-3}$ \msun{} \citep{kitaura2006} compared to their more massive FeCC-SN counterparts ($\approx$10$^{51}$ erg, $\sim$ 5 $\times$ 10$^{-2}$ \msun).
The envelopes of super-AGB star progenitors of EC-SNe may be either C-rich or O-rich prior to explosion, with dredge-out events being able to form carbon stars, albeit only for a short time before efficient HBB would burn this C to N resulting in C/O $<$ 1.

Owing to the substantial difference in envelope configurations and mass loss histories that are possible prior to an EC-SN, there is potential for a wide variation in observable events.  Hence many different classes of SNe have been suggested to originate from EC-SNe:

\begin{enumerate}[(i)]

\item the sub-luminous class of II-P including SN2003gd, SN2005cs, SN2008bk, SN2009md \citep{kitaura2006}
\item the so-called ``supernova impostors'' of SN2008S \citep{bot09,pumo2009}, SN2008ha \citep{fol09,val09} and NGC300-OT \citep{pri09};
\item a subset of the IIn-P class which includes SN1054 
\citep[the Crab nebula progenitor;][]{dav82,nom82,tom13} SN1994W, SN2009kn and SN2011ht \citep{smi13}
\end{enumerate}

Due to their low explosion energy, it had been assumed that EC-SNe may 
also be of low luminosity. Hence they have been associated with the 
sub-luminous members of the II-P SNe categrory \cite[e.g.][]{kitaura2006}. 
This association has fallen out of favour recently
as a result of modelling of EC-SN lightcurves, which found 
them to be as bright as typical Type II-P SNe \citep{tom13,mor14} 
due to their large progenitor radius and small envelope mass.
Another problem for this scenario comes from archival pre-SN proprieties of SN 2005cs. These 
showed that the progenitor 
had a luminosity far too low to be consistent with models in the post-SDU/dredge-out phase
of super-AGB evolution \citep{eld07}. 
Probably the most accepted model is that these sub-luminous SN have slightly more massive progenitors $\sim$ 10--15\msun{} \cite[i.e][]{spiro2014}.
Whilst there are some very promising candidate EC-SN, in particular SN1054 and 
SN2008S, as yet there has been no definitive confirmed super-AGB EC-SN events.


Based on observations of Type II-P SNe which also had pre-SN progenitor imaging, the review by \cite{sma09a} estimated the lower initial mass for CC-SN explosions to be $8.5^{+1}_{-1.5}$ \msun{}. However with an increased data set, this value was revised to $9.5^{+0.5}_{-2}$\msun{} or $10^{+0.5}_{-1}$\msun{} depending on choice of stellar models \citep{smartt2015}. These values for $M_{\rm{n}}$ are in reasonable agreement with the theoretical stellar evolutionary calculations discussed in Section~\ref{critmass} which take into account a moderate amount of overshooting. Under the assumption of the minimum initial mass for a CC-SN to be 8 \msun, the recent SUDARE supernova survey \citep{bot17} found the expected CC-SN rate to be higher than the value deduced from observations by about a factor of two. A possible remedy to this discrepancy is if the minimum mass for a CC-SN was increased to 10 \msun.

From an explosive nucleosynthesis perspective super-AGB stars that undergo an EC-SN may produce a wide variety of elements from Zn to Zr \citep{wanajo2011}, and isotopes \chem{48}Ca and \chem{60}Fe \citep{wanajo2013a,wanajo2013b}. Based on their nucleosynthesis calculations of \iso{86}Kr, \cite{wanajo2011} suggested the frequency of EC-SNe relative to all CC-SNe must be $\sim$ 4$\%$, assuming that the vast majority of this isotope in the solar system originates from EC-SNe. However \cite{wanajo2017} found that the nucleosynthesis from core collapse models of either O-Ne-Mg cores or the lowest mass Fe cores (dubbed EC-SNe like) is almost identical. Owing to this inability to distinguish resultant yields they suggest the EC-SN and low-mass FeCC-SN channel should be about 0.5--1 \msun{} wide in initial mass, if these events are the main source of elements Zn to Zr in the Galaxy.   

Recent determinations of the neutron star mass distribution have suggested that it may be bimodal \citep{sch10,kni11,val11}, with the lower mass peak suspected to arise from EC-SNe \citep{van04,podsi2004}. Due to the steep density gradient at the core edge, it is expected that EC-SNe result in neutron star masses very close to that of the original ONeMg core and with relatively small velocities \citep{podsi05,rad17}. However, recent low mass FeCC-SN and EC-SN progenitor models show smaller differences in the density gradients than in previous calculations \cite[e.g. see Figure 1 in][]{mueller2016} and due to this have far more similar explosion properties than in previous works, with these lowest mass FeCC-SNe dubbed ``EC-SN like''. Owing to their similarity in explosion properties the resulting neutron star masses may not be as clearly distinguishable as was thought, making it harder to separate neutron stars from the lowest mass FeCC-SN  and EC-Se. 

As described in Sect.~\ref{sec:binary}, binary evolution opens new evolutionary channels to EC-SNe. In the scenario that involves the merger of two WDs, the signature is expected to be that of a SN of Type Ib \citep{Nomoto_Hashimnoto1987}. If we consider a CV-like configuration of an accreting ONe WD, the explosion properties depend on the mass transfer rate. If accretion is slow, helium will build up at the surface of the WD and eventually ignite off-center. The detonation will then produce observational signatures similar to those of a SNIa \citep{Marquardt2015}. 
In the other class of models where the progenitor loses its H-rich envelope and becomes a He star, the EC-SN explosion would be observed as a SNIb or Ic depending of the mass and composition of the envelope surrounding the collapsing ONe core \citep{tauris2015}. According to \cite{Moriya2016}, the explosion of stripped-envelope EC-SNe could give rise to fast evolving transients and the newly-formed neutron star would have a low kick velocity.

  \subsection{Critical Uncertainties}

  \subsubsection{Mass loss}

  The very large uncertainty in the mass loss rates for super-AGB stars, particularly at low metallicity, is a major impediment to determining the final fates of this class of star. Whilst there are observational studies examining the impact of metallicity on the mass loss rates of AGB and red super-giant  stars  \cite[e.g.][]{goldman2017}, these works can only probe the relatively metal rich populations, because the low metallicity super-AGB stars are
  long since dead. 
However, there are valiant efforts underway in theoretical modelling of AGB star atmospheres and circumstellar envelopes to derive a predictive theory of mass loss for (super-)AGB stars (see \citealt{hof16} and reference therein). 
 In addition to the importance of determining the ``standard''  mass loss from super-AGB stars, there may also be mass expulsions from  either the Fe-peak instability \citep{lau2012} or global oscillation of shell-H ingestion  \citep[GOSH, see ][]{herwig2014}. Both of these phenomena may have a substantial impact on the frequency of EC-SNe, as well as potentially modify super-AGB nucleosynthesis and need further examination. Based on observations of evolved massive AGB stars \cite{dev14,dev15} have suggested the need for a ``hyper''-wind of 10$^{-3}$-10$^{-2}$ \msun{} yr$^{-1}$ to account for the very short superwind duration and lack of extended structure \citep{cox12}. Could this hyperwind be as a result of an Fe peak instability?

 \subsubsection{Convection}

 Our current theory of convection, in particular the treatment of convective boundaries, lies at the heart of the majority of the uncertainties in this mass range.
 Arguably the most important factor that effects the mass boundaries $M_{\rm{up}}$ and $M_{\rm{mas}}$ is the treatment of convective borders, in particular during core He burning. Multidimensional hydrodynamical simulations 
 of convection are key to understanding this mass range.  A new theory of convection, such as the 321D 
 theory under development \citep{arnett2015,campbell2016}, would help to 
 constrain the WD/SN boundary . 

Is convective boundary mixing efficient enough to stall C, and/or Ne burning i.e. Do CO-Ne WDs or FMS exist? While the study of \cite{lec16} suggested the C-burning would most likely not be stalled by CBM, the results from C burning flame propagation cannot simply be applied to that of neon-oxygen burning flames to rule out the formation of FMS. This is due to the structure between C and Ne flames being quite different with Ne burning flames far thinner and faster moving \cite[e.g.][]{timmes1994}. Multi dimensional calculations of Ne flames are crucial to determine if FMS exist.

What is the impact of convective boundary mixing on the occurrence and efficiency of third dredge-up? 
Are super-AGB stars the site of the i-process heavy element production from during proton ingestion TPs \citep{jones2016a}? 

\subsection{Research Questions}

There are a number of investigations that could be done in the near future, and which would dramatically advance this field. We list a selection of these here.

\begin{itemize}
\item an analysis of red super-giant surveys to possibly identify super-AGB;
\item further analysis of GOSH events, to determine their implications for super-AGB stars;
\item the problem of convergence and the Fe opacity peak; this will require hydrodynamical models but could yield significant advances in our understanding of the masses and progenitors of various SNe;
\item hydrodynamical multi-dimensional flame propagation calculations, especially for the Ne flame;
\item reevaluation/measurement of the critical nuclear reaction rates: \iso{23}Na + p, \iso{12}C+\iso{12}C;
\item details of dust production from super-AGB and massive AGB stars;
\item proton ingestion and the resulting i-process neutron captures; this may have implications well beyond super-AGB stars.
\end{itemize}

\subsection{Conclusion}

The last decade has seen real advances in the study of
the lives and deaths of super-AGB stars. There are now fully detailed 
interior stellar evolutionary models along the entire TP-SAGB for a wide range of 
metallicities from $Z=0.04$ to $Z=10^{-5}$ (see companion review by Gil Pons, 2017). 
Furthermore super-AGB star nucleosynthesis predictions 
for both light and heavy elements are now available for
some compositions, and the composition range is being extended all the time. 
The first detailed dust yields are also now available. 

The study of EC-SNe has also flourished in recent years. 
The production of a new generation of single star EC-SN 
progenitor models is an important update on the previous models from the early 1980s.
There are now detailed single and multi-dimensional simulations of EC-SNe (and EC-SN like) explosions as well as detailed nucleosynthetic calculations from EC-SNe. In addition, recent works have calculated synthetic light-curves of EC-SNe within super-AGB star envelopes.
There have also been important advances in the role of binary stars in producing EC-SNe. 

All of these above results have produced predictions that can and are being tested against observation to further constrain these stars on the low/high mass star boundary.

Nevertheless, some of the most important questions remain, for example the recent multi-dimensional simulations of O-deflagration in ONe cores has reopened the debate of the final fate of an EC-SN on whether they explode by oxygen deflagration, or collapse by electron captures 
to form a neutron star.

We have highlighted/identified areas that need attention, which may help us finally answer some of these important questions.

\begin{acknowledgements}
This work was supported in part by the National Science Foundation under Grant No. PHY-1430152 (JINA Center for the Evolution of the Elements).
CD acknowledges support from the Lendulet-2014 Programme of the Hungarian Academy of Sciences. CD would like to thank S.W Campbell and S. Jones for interesting discussions. LS is a senior FNRS researcher.
\end{acknowledgements}

\bibliographystyle{pasa-mnras}
\bibliography{pasa-references}

\end{document}